\documentclass[preprint]{elsarticle}

\usepackage{adforn} 

\usepackage[T1]{fontenc}
\usepackage{esvect}

\usepackage[dvipsnames]{xcolor}
\usepackage{graphicx}
\usepackage{tikz}
\usetikzlibrary{arrows,decorations.pathmorphing,backgrounds,positioning,fit,petri,automata,shadows,calendar,mindmap,decorations.markings,calc}
\tikzset{->-/.style={decoration={markings, mark=at position .5 with {\arrow{>}}},postaction={decorate}}}
\tikzset{-<-/.style={decoration={markings, mark=at position .5 with {\arrow{<}}},postaction={decorate}}}
\usepackage{cancel}
\usepackage{pdflscape}
\usepackage{float}
\usepackage{./StyleFilesMacros/axodraw}
\usepackage{lineno}
\usepackage{enumitem}
\setlist{nolistsep} 
\usepackage{heppennames2}
\usepackage{tabularx}
\usepackage{cernunits}
\usepackage{./StyleFilesMacros/lhchiggs}

\usepackage{./StyleFilesMacros/hepparticles}
\usepackage{multirow}

\usepackage{fancyhdr}
\usepackage{tikz}

\newcommand{\cref}[1]{Chapter~\ref{c.#1}}

\newcommand{\nc}{\newcommand}
\nc{\gev}{\;\mathrm{GeV}}

\usepackage{hypernat}
\usepackage[toc,page]{appendix}
\usepackage{booktabs}
\usepackage{bigdelim}
\usepackage[hyperfootnotes=true,bookmarks=false,hypertexnames=true,pagebackref=true,linktocpage=false]{hyperref}

\renewcommand*\backref[1]{\ifx#1\relax \else (#1) \fi}

\usepackage{chngcntr}
\usepackage[subfigure]{tocloft} 
\usepackage{subfigure} 
\usepackage{wrapfig}
\usepackage{afterpage}

\usepackage[export]{adjustbox} 

\usepackage[export]{adjustbox}

\title{Vector Boson Scattering: recent experimental and theory developments}

\usepackage{layout}

\def\be{\begin{equation}}
\def\ee{\end{equation}}

\providecommand{\MWOS}{}
\renewcommand{\MWOS}{\ensuremath{M_\PW^\text{OS}}\xspace}
\providecommand{\MZOS}{}
\renewcommand{\MZOS}{\ensuremath{M_\PZ^\text{OS}}\xspace}
\providecommand{\GZOS}{}
\renewcommand{\GZOS}{\ensuremath{\Gamma_\PZ^\text{OS}}\xspace}
\providecommand{\GWOS}{}
\renewcommand{\GWOS}{\ensuremath{\Gamma_\PW^\text{OS}}\xspace}

\providecommand{\MVOS}{}
\renewcommand{\MVOS}{\ensuremath{M_V^\text{OS}}\xspace}%
\providecommand{\GVOS}{}
\renewcommand{\GVOS}{\ensuremath{\Gamma_V^\text{OS}}\xspace}%

\providecommand{\GeV}{}
\renewcommand{\GeV}{\ensuremath{\,\text{GeV}}\xspace}
\providecommand{\TeV}{}
\renewcommand{\TeV}{\ensuremath{\,\text{TeV}}\xspace}

\providecommand{\alphas}{}
\renewcommand{\alphas}{\ensuremath{\alpha_\text{s}}\xspace}

\providecommand{\ptsub}{}
\renewcommand{\ptsub}[1]{\ensuremath{p_{\text{T},#1}}\xspace}

\providecommand{\etsub}{}
\renewcommand{\etsub}[1]{\ensuremath{E_{\text{T},#1}}\xspace}

\newcolumntype{.}{D{.}{.}{-1}}
\newcolumntype{d}[1]{D{.}{.}{#1}}

\renewcommand{\vec}[1]{\mathbf{#1}}

\newlength{\width}
\newlength{\height}

\def\draftdate{\relax}
\def\mda{\relax}
\def\mua{\relax}
\def\mla{\relax}
\def\draft{
\def\thtystars{******************************}
\def\sixtystars{\thtystars\thtystars}
\typeout{}
\typeout{\sixtystars**}
\typeout{* Draft mode!
         For final version remove \protect\draft\space in source file *}
\typeout{\sixtystars**}
\typeout{}
\def\draftdate{\today}
\def\mua{\marginpar[\boldmath\hfil$\uparrow$]%
                   {\boldmath$\uparrow$\hfil}\color{black}%
                    \typeout{marginpar: $\uparrow$}\ignorespaces}
\def\mda{\color{red}\marginpar[\boldmath\hfil$\downarrow$]%
                   {\boldmath$\downarrow$\hfil}%
                    \typeout{marginpar: $\downarrow$}\ignorespaces}
\def\mla{\marginpar[\boldmath\hfil$\rightarrow$]%
                   {\boldmath$\leftarrow $\hfil}%
                    \typeout{marginpar: $\leftrightarrow$}\ignorespaces}
\def\Mua{\marginpar[\boldmath\hfil$\Uparrow$]%
                   {\boldmath$\Uparrow$\hfil}\color{black}%
                    \typeout{marginpar: $\uparrow$}\ignorespaces}
\def\Mda{\color{red}\marginpar[\boldmath\hfil$\Downarrow$]%
                   {\boldmath$\Downarrow$\hfil}%
                    \typeout{marginpar: $\downarrow$}\ignorespaces}
\def\Mla{\marginpar[\boldmath\hfil\textcolor{red}{$\Rightarrow$}]%
                   {\boldmath\textcolor{red}{$\Leftarrow $}\hfil}%
                    \typeout{marginpar: $\leftrightarrow$}\ignorespaces}
\overfullrule 5pt
\oddsidemargin 15mm
\marginparwidth 29mm
}

\def\be{\begin{equation}}
\def\ee{\end{equation}}

\providecommand{\Pj}{}
\renewcommand{\Pj}{\ensuremath{\text{j}}\xspace}
\providecommand{\MWOS}{}
\renewcommand{\MWOS}{\ensuremath{M_{\rm W}^{\text{OS}}}\xspace}
\providecommand{\MZOS}{}
\renewcommand{\MZOS}{\ensuremath{M_{\rm Z}^{\text{OS}}}\xspace}
\providecommand{\GZOS}{}
\renewcommand{\GZOS}{\ensuremath{\Gamma_{\rm Z}^{\text{OS}}}\xspace}
\providecommand{\GWOS}{}
\renewcommand{\GWOS}{\ensuremath{\Gamma_{\rm W}^{\text{OS}}}\xspace}

\providecommand{\MVOS}{}
\renewcommand{\MVOS}{\ensuremath{M_V^{\text{OS}}}\xspace}%
\providecommand{\GVOS}{}
\renewcommand{\GVOS}{\ensuremath{\Gamma_V^{\text{OS}}}\xspace}%

\providecommand{\GeV}{}
\renewcommand{\GeV}{\ensuremath{\,\text{GeV}}\xspace}
\providecommand{\TeV}{}
\renewcommand{\TeV}{\ensuremath{\,\text{TeV}}\xspace}

\providecommand{\ptsub}{}
\renewcommand{\ptsub}[1]{\ensuremath{p_{\text{T},#1}}\xspace}

\providecommand{\etsub}{}
\renewcommand{\etsub}[1]{\ensuremath{E_{\text{T},#1}}\xspace}

\newcolumntype{.}{D{.}{.}{-1}}
\newcolumntype{d}[1]{D{.}{.}{#1}}

\renewcommand{\vec}[1]{\mathbf{#1}}

\def\draftdate{\relax}
\def\mda{\relax}
\def\mua{\relax}
\def\mla{\relax}
\def\draft{
\def\thtystars{******************************}
\def\sixtystars{\thtystars\thtystars}
\typeout{}
\typeout{\sixtystars**}
\typeout{* Draft mode!
         For final version remove \protect\draft\space in source file *}
\typeout{\sixtystars**}
\typeout{}
\def\draftdate{\today}
\def\mua{\marginpar[\boldmath\hfil$\uparrow$]%
                   {\boldmath$\uparrow$\hfil}\color{black}%
                    \typeout{marginpar: $\uparrow$}\ignorespaces}
\def\mda{\color{red}\marginpar[\boldmath\hfil$\downarrow$]%
                   {\boldmath$\downarrow$\hfil}%
                    \typeout{marginpar: $\downarrow$}\ignorespaces}
\def\mla{\marginpar[\boldmath\hfil$\rightarrow$]%
                   {\boldmath$\leftarrow $\hfil}%
                    \typeout{marginpar: $\leftrightarrow$}\ignorespaces}
\def\Mua{\marginpar[\boldmath\hfil$\Uparrow$]%
                   {\boldmath$\Uparrow$\hfil}\color{black}%
                    \typeout{marginpar: $\uparrow$}\ignorespaces}
\def\Mda{\color{red}\marginpar[\boldmath\hfil$\Downarrow$]%
                   {\boldmath$\Downarrow$\hfil}%
                    \typeout{marginpar: $\downarrow$}\ignorespaces}
\def\Mla{\marginpar[\boldmath\hfil\textcolor{red}{$\Rightarrow$}]%
                   {\boldmath\textcolor{red}{$\Leftarrow $}\hfil}%
                    \typeout{marginpar: $\leftrightarrow$}\ignorespaces}
\overfullrule 5pt
\oddsidemargin 15mm
\marginparwidth 29mm
}

\author[TOI]{Ballestrero, Alessandro}
\author[TOR]{Bellan, Riccardo}
\author[WUR]{Biedermann, Benedikt}
\author[TUD]{Bittrich, Carsten}
\author[NIE]{Brivio, Ilaria}            
\author[NIK]{Bruni, Lucrezia Stella}    
\author[FIR]{Cardini, Andrea}           
\author[MIT]{Gomez-Ceballos, Guillelmo}
\author[LLR]{Charlot, Claude}     
\author[FIR]{Ciulli, Vitaliano}      

% ...
\address[FIR]{University and INFN, Firenze (IT)}
% ...
\address[LLR]{LLR, \'Ecole polytechnique, CNRS/IN2P3, Universit\'e Paris-Saclay (FR)}
% ...
\address[MIT]{Massachusetts Institute of Technology, Cambridge (US)}
% ...
\address[NIE]{Niels Bohr International Academy and Discovery Center, Niels Bohr Institute, Copenhagen University (DK)}
\address[NIK]{Nikhef National institute for subatomic physics (NL)}
% ...
\address[TOI]{INFN Torino (IT)}
\address[TOR]{University and INFN Torino (IT)}
% ...
\address[TUD]{Technische Universitaet Dresden (DE)}
% ...
\address[WUR]{University of W\"urzburg (DE)}

\begin{document}



  




\begin{abstract}
This document summarises the talks and discussions happened during the VBSCan Split17 workshop,
the first general meeting of the VBSCan COST Action network.
This collaboration is aiming at a consistent and coordinated study of vector-boson scattering
from the phenomenological and experimental point of view, 
for the best exploitation of the data that will be delivered by 
existing and future particle colliders.
\end{abstract}

\maketitle

%
%





\cleardoublepage
\phantomsection
\addcontentsline{toc}{part}{Introduction}

\section{Introduction}
\label{chapter:intro}

In the past years, the study of the vector-boson scattering (VBS) 
 attracted lots of interest in the theory and experimental communities
(see \emph{e.g.}\ References~\cite{Rauch:2016pai, Green:2016trm} for recent reviews).
If the Standard Model (SM) is a partial description of Nature 
and its completion happens at energies which are experimentally unreachable,
the cross section of VBS processes could increase substantially between the Higgs boson mass
and the scale at which new physics mechanisms intervene to restore the unitarity of the process.
The study of VBS offers then the unique opportunity to seize
this scenario of ``delayed unitarity cancellations"
even if the energy scale at which the processes beyond the SM (BSM) enter into play
goes beyond our experimental reach, 
either today or in the next future.
This measurement, though, presents several challenges,
both on the theory side and on the experimental one.
For example difficulties arise, at hadron colliders,
because of the complexity of the final state already at tree level,
where six fermions arise from the interaction vertex of the initial state partons.
Four of them are expected to be the product of vector-boson disintegrations,
while two to be remnants of the colliding protons.
In this complex environment, 
the exact calculation of the process needs to be kept as reference to validate any approximations
performed to understand the main features of the process.
Effective field theory parameterisations of BSM theories 
have to cope with the multiplex final state,
and higher order corrections are not easily calculated either.
Experimentally,
several processes give rise to signatures similar to the VBS ones,
and the events typically span a large angle in rapidity,
involving the entire particle detectors in the measurement.
Therefore, algorithms need to be optimised to reject backgrounds at best,
involving all the sub-elements of each measuring apparatus,
and featuring the most advanced data analysis techniques.

Only a coherent action in the experimental and theoretical community
will grant that all these challenges will be met
and that the data delivered by current and future particle colliders
will be exploited at best.
The VBSCan COST Action is a four-year project,
funded by the Horizon 2020 Framework Programme of the European Union,
aiming at a consistent and coordinated study of VBS
from the phenomenological and experimental points of view, 
gathering all the interested parties in the high-energy physics community,
together with experts of data mining techniques.

The community is organised in five working groups,
three of which are focussing on the scientific aspects of the collaboration.
One is dedicated to the theoretical understanding of the VBS process (WG1),
which targets a detailed description of the signal 
and relative backgrounds in the SM, 
as well as effective field theory (EFT) modelling of BSM effects.
A second one focuses on analysis techniques (WG2),
studying the definition of data analysis protocols and performances 
to maximise the significance of the VBS analyses at hadron colliders, 
promoting the communication between theory and experiments.
A third one fosters the optimal deployment of the studies 
in the hadron collider experiments data analyses (WG3).
The remaining two working groups address the knowledge exchange and cross-activities (WG4),
and the implementation of the COST inclusiveness policies (WG5),
respectively.

The Network is composed of theoretical and experimental physicists from both the ATLAS and CMS experiments,
as well as data analysis experts and industrial partners.
The first general meeting marked the start of the activities.
It happened at the end of June 2017 in Split~\cite{kickoff}
and was dedicated to reviews of the data analysis status of the art,
as well as of the theoretical and experimental instruments
relevant for VBS studies.
This report contains
a summary of the talks presented
\footnote{All the presentations can be found at \url{https://indico.cern.ch/event/629638/}.}
divided into sections corresponding to the Network working groups.
A review of the theoretical understanding
and the parameterisation of the impact of new physics in the VBS domain
through effective field theory expansion
is given in Section 1;
an outline of the existing experimental results at the time of the Split workshop, 
and the overview of future prespectives are presented in Section 2; 
we conclude in Section 3 with an overview of the existing techniques for the identification of jets
in the ATLAS and CMS experiments, 
which are the most complicated physics objects to be dealt with in VBS
and, therefore, the main focus of WG3.





\section{WG1: theoretical understanding}

\subsection{Complete NLO corrections to ${\rm W^+ W^+}$ scattering}\footnote{speaker: M. Pellen}

The first VBS process that has been observed during the run~I of the LHC 
is the same-sign WW production~\cite{Aad:2014zda,Aaboud:2016ffv,Khachatryan:2014sta}.
This observation has already been confirmed 
by a measurement of the CMS collaboration at the 13 TeV run~II~\cite{CMS:2017adb}.
In view of the growing mole of data which will be collected by the experiments, 
and of the consequent reduction of the uncertainties affecting these measurements, 
precise theoretical predictions become necessary.

In that respect 
next-to-leading order (NLO) QCD and electroweak (EW) corrections to such signatures 
should be computed.
So far, 
NLO computations have focused 
on NLO QCD corrections to the VBS process~\cite{Jager:2009xx,Jager:2011ms,Denner:2012dz,Rauch:2016pai} 
and its QCD-induced irreducible background process~\cite{Melia:2010bm,Melia:2011gk,Campanario:2013gea,Baglio:2014uba,Rauch:2016pai}.
No NLO EW corrections had been computed 
and the NLO QCD computations relied on the so-called VBS approximation.
In Reference~\cite{Biedermann:2017bss}, 
for the first time, 
all leading order (LO) and NLO contributions 
to the full ${\rm p}{\rm p}\to\mu^+\nu_\mu{\rm e}^+\nu_{\rm e}{\rm j}{\rm j}$ process 
have been reported\footnote{The $\mathcal{O}{(\alpha^{7})}$ corrections
were computed previously in Reference~\cite{Biedermann:2016yds}.}.
As the full amplitudes are used, 
this amounts to computing three LO contributions and four NLO contributions.
At LO, 
the three contributions are the EW process [order $\mathcal{O}{(\alpha^{6})}$], 
its QCD-induced counterpart [order $\mathcal{O}{(\alpha_{\rm s}^2\alpha^{4})}$] 
as well as the interference [order $\mathcal{O}{(\alpha_{\rm s}\alpha^{5})}$].
Due to the VBS event selection applied to the final state, 
the full process is dominated by the purely EW contribution (see Table~\ref{table:LOVBS}).
This EW contribution features the proper VBS diagrams but also background diagrams where, 
for example, 
the W bosons are simply radiated off the quark lines.

\begin{table}
\begin{center}
\begin{tabular}{l|c|c|c|c}
Order & $\mathcal{O}{(\alpha^{6})}$ & $\mathcal{O}{(\alpha_{\rm s}\alpha^{5})}$ & $\mathcal{O}{(\alpha_{\rm s}^2\alpha^{4})}$ & Sum \\
\hline
\hline
${\sigma_{\mathrm{LO}}}$ [fb] 
& $1.4178(2)$
& $0.04815(2)$
& $0.17229(5)$
& $1.6383(2)$ \\
\end{tabular}
\end{center}
\caption{
Fiducial cross section from Reference~\cite{Biedermann:2017bss} at $\sqrt{s}$=13~TeV at LO for the process ${\rm p}{\rm p}\to\mu^+\nu_\mu{\rm e}^+\nu_{\rm e}{\rm j}{\rm j}$, at
orders  $\mathcal{O}{(\alpha^{6})}$, $\mathcal{O}{(\alpha_{\rm s}\alpha^{5})}$, and $\mathcal{O}{(\alpha_{\rm s}^2\alpha^{4})}$.
The sum of all the LO contributions is in the last column and all contributions are expressed in femtobarn. 
The statistical uncertainty from the Monte Carlo integration on the last digit is given in parenthesis.}
\label{table:LOVBS}
\end{table}

At NLO, 
the four contributions arise at the orders $\mathcal{O}{(\alpha^{7})}$, 
$\mathcal{O}{(\alpha_{\rm s}\alpha^{6})}$, $\mathcal{O}{(\alpha_{\rm s}^{2}\alpha^{5})}$, 
and $\mathcal{O}{(\alpha_{\rm s}^{3}\alpha^{4})}$. 
An interesting feature is that 
the orders $\mathcal{O}{(\alpha_{\rm s}\alpha^{6})}$ and $\mathcal{O}{(\alpha_{\rm s}^{2}\alpha^{5})}$ 
receive both EW and QCD corrections.
Thus, 
at NLO (as opposed to LO) it is not possible to strictly distinguish the EW process from the QCD-induced one.
As it can be seen from Table~\ref{table:NLOVBS}, 
at the level of the fiducial cross section, 
the largest corrections are the ones of order $\mathcal{O}{(\alpha^{7})}$.
These are the NLO EW corrections to the EW processes.

\begin{table}
\begin{center}
\begin{tabular}{l|c|c|c|c|c}
Order & $\mathcal{O}{(\alpha^{7})}$ & $\mathcal{O}{(\alpha_{\rm s}\alpha^{6})}$ & $\mathcal{O}{(\alpha_{\rm s}^{2}\alpha^{5})}$ & $\mathcal{O}{(\alpha_{\rm s}^{3}\alpha^{4})}$ & Sum \\
\hline
\hline 
${\delta \sigma_{\mathrm{NLO}}}$ [fb] 
& $-0.2169(3)$ 
& $-0.0568(5)$
& $-0.00032(13)$
& $-0.0063(4)$ 
& $-0.2804(7)$ \\
\hline
$\delta \sigma_{\mathrm{NLO}}/\sigma_{\rm LO}$ [\%] & $-13.2$ & $-3.5$ & $0.0$ & $-0.4$ & $-17.1$ \\
\end{tabular}
\end{center}
\caption{
NLO corrections from Reference~\cite{Biedermann:2017bss} for the process ${\rm p}{\rm p}\to\mu^+\nu_\mu{\rm e}^+\nu_{\rm e}{\rm j}{\rm j}$ at the orders 
$\mathcal{O}{(\alpha^{7})}$, $\mathcal{O}{(\alpha_{\rm s}\alpha^{6})}$, $\mathcal{O}{(\alpha_{\rm s}^{2}\alpha^{5})}$, and $\mathcal{O}{(\alpha_{\rm s}^{3}\alpha^{4})}$.
The sum of all the NLO contributions is in the last column.
The contribution $\delta\sigma_{\mathrm{NLO}}$ corresponds to the absolute correction while $\delta \sigma_{\mathrm{NLO}}/\sigma_{\rm LO}$ gives the relative correction normalised to the sum of all LO contributions.
The absolute contributions are expressed in femtobarn while the relative ones are expressed in per cent.
The statistical uncertainty from the Monte Carlo integration on the last digit is given in parenthesis.}
\label{table:NLOVBS}
\end{table}

\begin{figure}
\includegraphics[width=.5\textwidth,trim={2.2cm 0 0.15cm 0},clip=true]{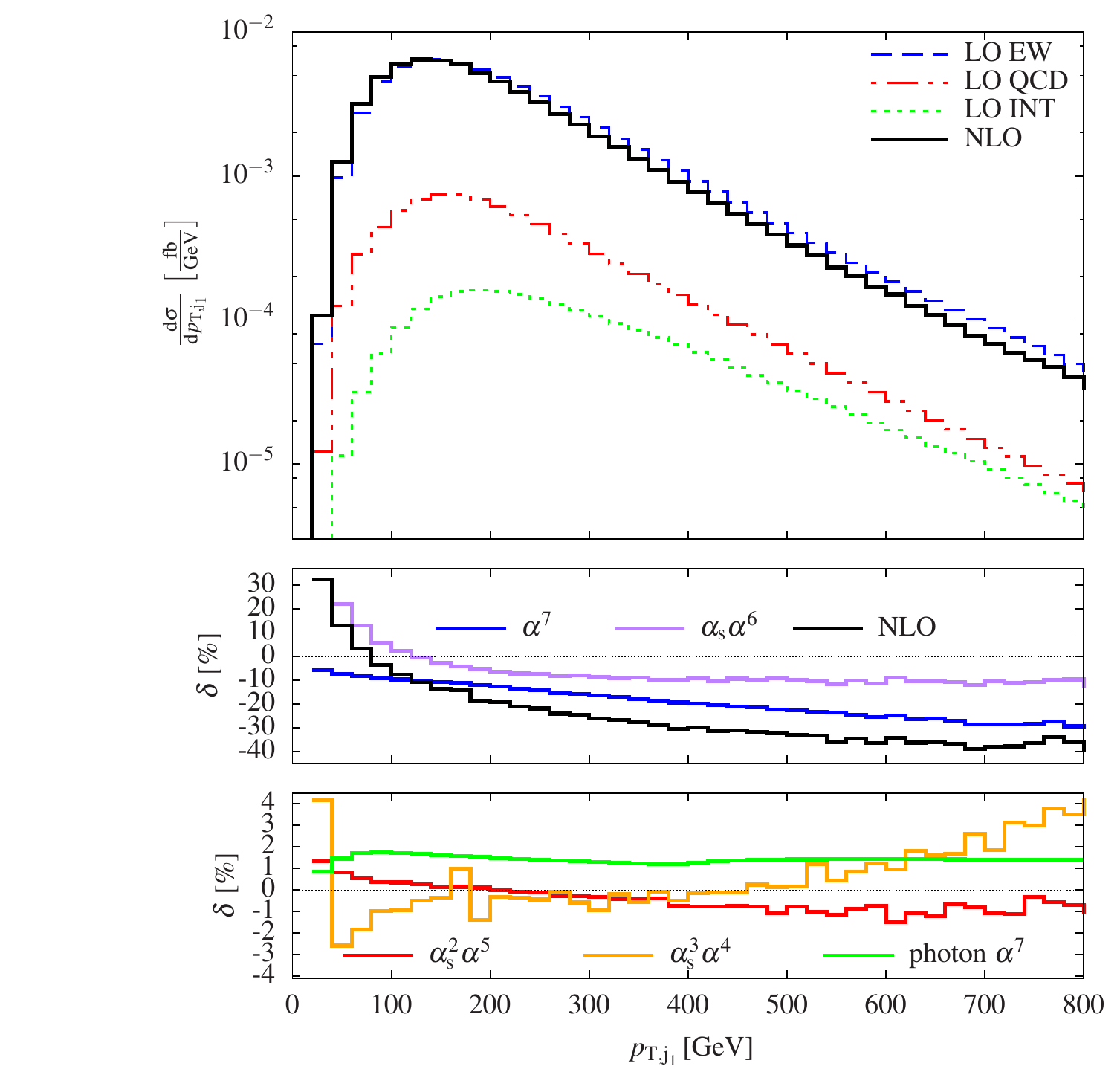}
\includegraphics[width=.5\textwidth,trim={2.2cm 0 0.15cm 0},clip=true]{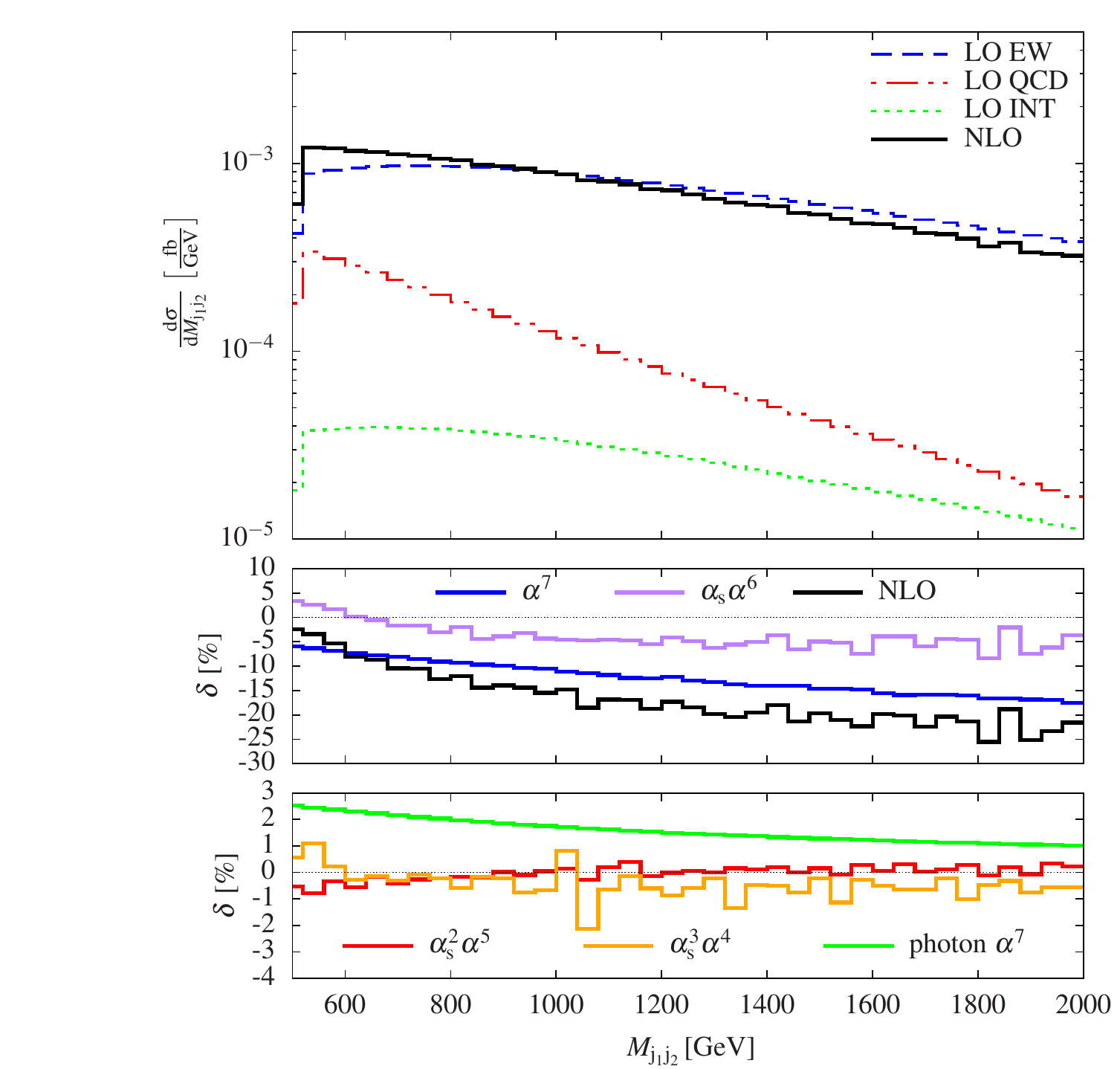}
\caption{Differential distributions at $\sqrt{s}$=13~TeV from Reference~\cite{Biedermann:2017bss} for ${\rm p}{\rm p}\to\mu^+\nu_\mu{\rm e}^+\nu_{\rm e}{\rm j}{\rm j}$:
transverse momentum for the hardest jet~(left) and invariant mass for the two leading jets~(right).
The two lower panels show the relative NLO corrections with respect to the full LO in per cent,
defined as $\delta_i = \delta \sigma_{i} / \sum \sigma_{\text{LO}}$, 
where $i=\mathcal{O}{(\alpha^{7})},\mathcal{O}{(\alpha_{\rm s}\alpha^{6})},\mathcal{O}{(\alpha_{\rm s}^2\alpha^{5})},\mathcal{O}{(\alpha_{\rm s}^3\alpha^{4})}$.
In addition, the NLO photon-induced contributions of order $\mathcal{O}{(\alpha^{7})}$ is provided separately.}
\label{fig:VBSALL}
\end{figure}

This is also reflected 
in two differential distributions in the transverse momentum for the hardest jet 
and invariant mass for the two leading jets in Figure~\ref{fig:VBSALL}.
At LO, 
the QCD-induced process as well as the interference 
are rather suppressed due to the typical VBS event selection 
(as for the fiducial cross section).
This is exemplified in the invariant mass of the two leading jets where, 
at high invariant mass, 
the QCD-induced background becomes negligible.
At NLO, 
the bulk of the corrections originates from the large EW corrections 
to the EW process at order $\mathcal{O}{(\alpha^{7})}$.
In particular, 
they display the typical behaviour of Sudakov logarithms that grow large in the high-energy limit.
Note that the contribution from initial-state photon is also shown 
but is not taken into account in the definition of the NLO predictions.
This contribution is rather small and relatively constant in shape over the whole range studied here.
Finally, 
as at NLO it is not possible to isolate the EW production from its irreducible backgrounds, 
a global measurement of the full process ${\rm p}{\rm p}\to\mu^+\nu_\mu{\rm e}^+\nu_{\rm e}{\rm j}{\rm j}$ 
with all components is desirable.

\begin{figure}
\includegraphics[width=.5\textwidth,trim={2.2cm 0 0.2cm 0},clip=true]{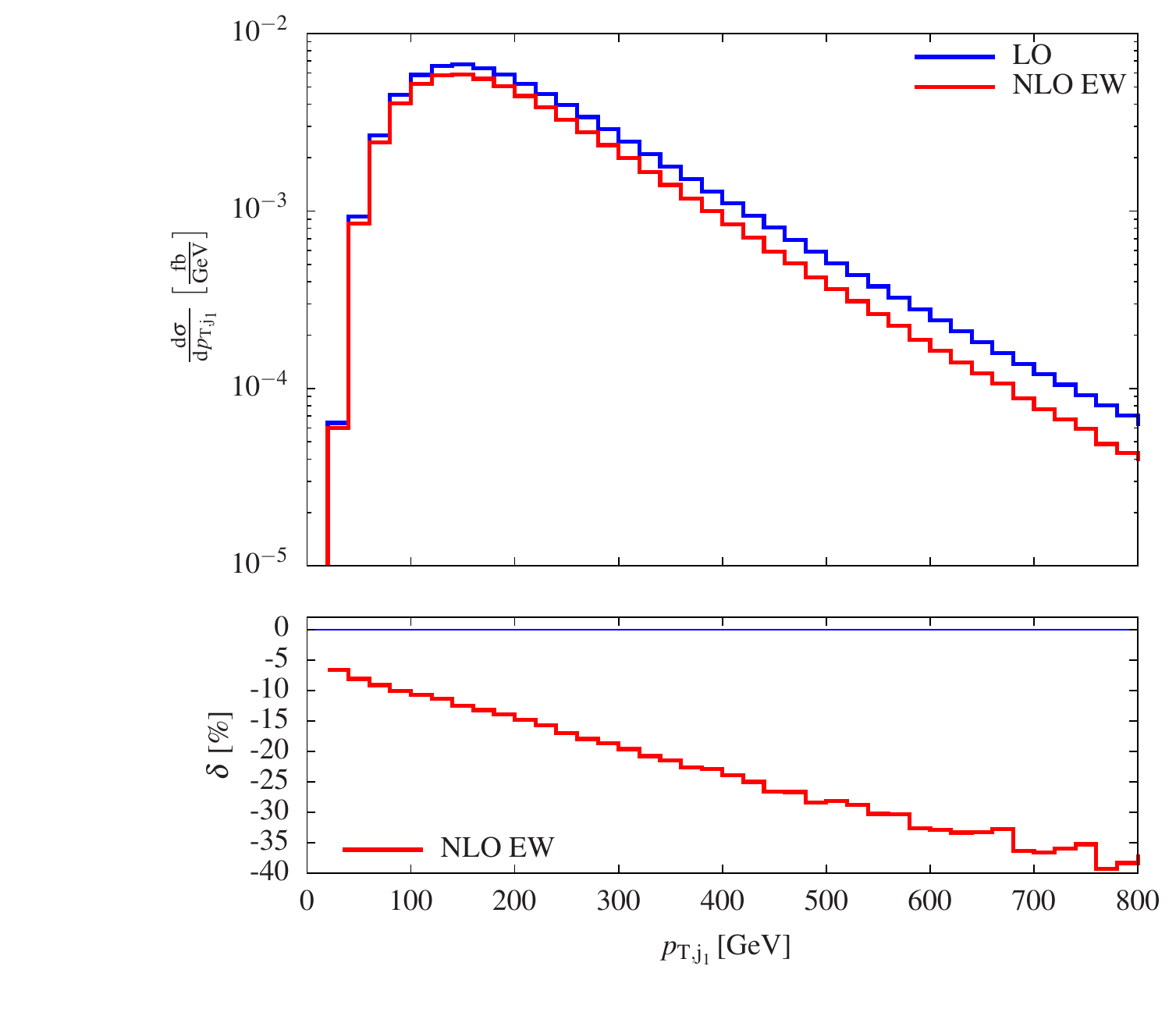}
\includegraphics[width=.5\textwidth,trim={2.2cm 0 0.2cm 0},clip=true]{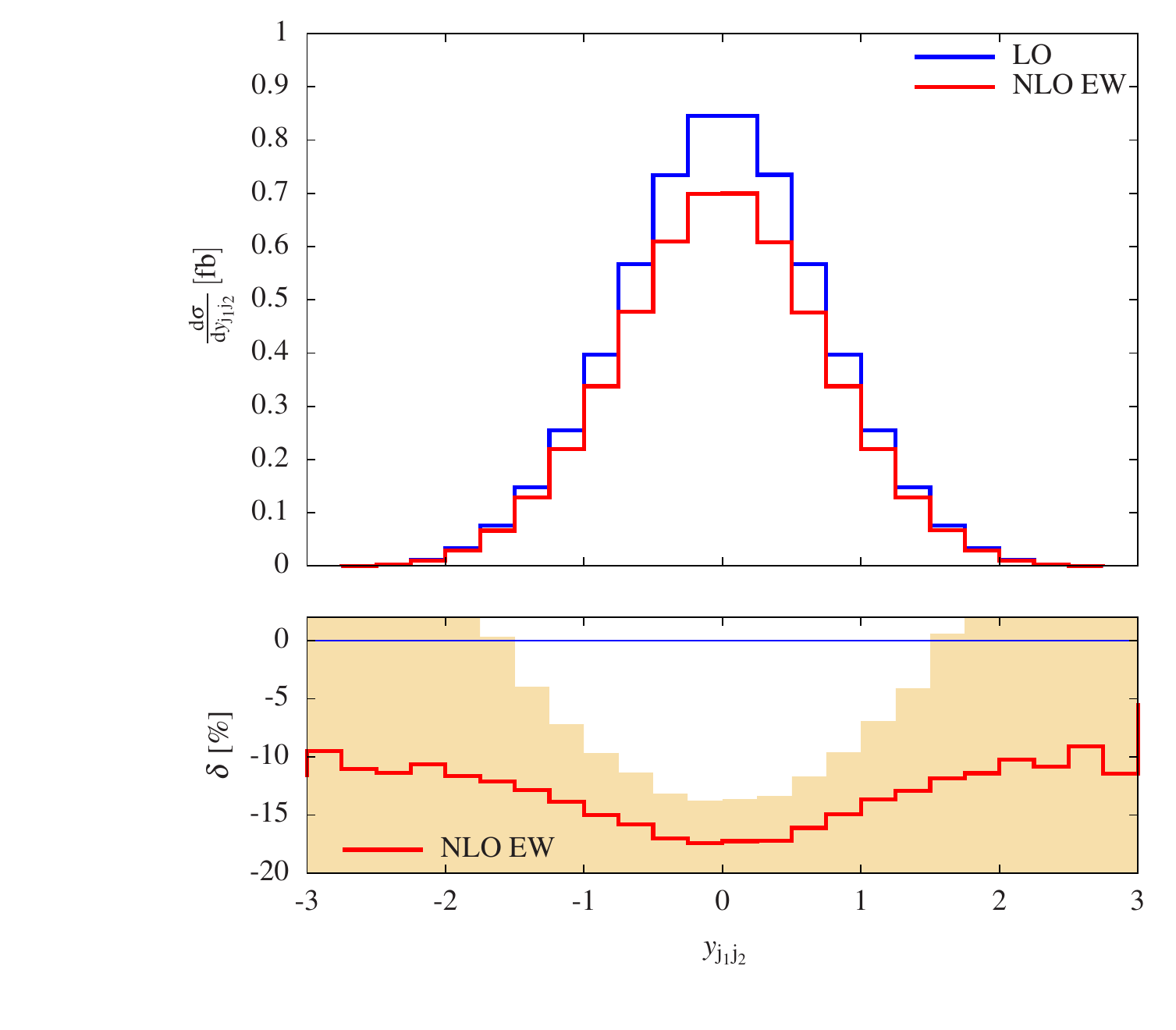}
\caption{Differential distributions from Reference~\cite{Biedermann:2016yds} 
         for ${\rm p}{\rm p}\to\mu^+\nu_\mu{\rm e}^+\nu_{\rm e}{\rm j}{\rm j}$ including NLO EW corrections 
         (upper panel) and relative NLO EW corrections (lower panel) at $\sqrt{s}$=13~TeV.
         Left plot: Transverse momentum of the most energetic jet.
         Right plot: Rapidity distribution of the leading jet pair.
         The yellow band describes the expected statistical experimental uncertainty 
         for a high-luminosity LHC collecting $3000~{\rm fb}^{-1}$ 
         and represents a relative variation of $\pm 1/\sqrt{N_{\rm obs}}$, 
         where $N_{\rm obs}$ is the number of observed events in each bin.}
\label{fig:VBSEW}
\end{figure}

We conclude by stressing that usually EW corrections are particularly sizeable 
only in phase space regions which are dominated by large scales and hence are 
suppressed.
Therefore the impact at the level of the total fiducial cross section is usually rather limited.
This is not the case here where the corrections are already large at the level of the cross section 
and reach $-17.1\%$ \cite{Biedermann:2016yds}.
The origin of these large EW corrections are virtual corrections 
and in particular the ones corresponding 
to the insertion of massive vector particles in the scattering process \cite{Biedermann:2016yds}.
Hence, large NLO EW corrections are an intrinsic feature of VBS at the LHC.
As the EW corrections are particularly large, 
it might be possible to measure them at a high luminosity LHC, 
hence probing the EW sector of the SM to very high precision.
This is illustrated on the right-hand side of Figure~\ref{fig:VBSEW} 
where the band represents the estimated statistical error for a high-luminosity LHC collecting $3000~{\rm fb}^{-1}$.

\subsection{Monte Carlo comparisons for ${\rm W^+ W^+}$ scattering}\footnote{speaker: M. Zaro}
\label{sec:MCcomparison}

This talk presents some preliminary results of a study
which has appeared during the publication of these proceedings~\cite{Ballestrero:2018anz}.
In the last decade many codes capable of performing VBS simulations have appeared.
Within a network such as VBSCan it is therefore natural to perform 
a quantitative comparison of these programs, both to cross-validate results and to assess the impact of the different approximations which are used.
In fact, already at LO, when considering the process ${\rm p}{\rm p}\to\mu^+\nu_\mu{\rm e}^+\nu_{\rm e}{\rm j}{\rm j}$ at order $\mathcal O (\alpha^6)$, the various implementations of VBS simulations are different.
They differ, for example,
by the (non-)inclusion of diagrams with vector bosons in the $s$-channel or by the treatment of interferences between diagrams.
The reason of these differences is that, when typical signal cuts for VBS are imposed, these effects turn out to be small on rates and distributions.

In the comparison, the following codes are used: 
\begin{itemize}
 \item The program {\sc Bonsay}~\cite{Schwan:2017yy} (contact person: C. Schwan)
       consists of a general-purpose Monte Carlo integrator
       and matrix elements taken from several sources: Born matrix elements are
       adapted from the program {\sc Lusifer} \cite{Dittmaier:2002ap} for the partonic
       processes, real matrix elements are written by Marina Billoni, and virtual
       matrix elements by Stefan Dittmaier.
       One loop integrals are evaluated using the {\sc Collier} library
       \cite{Denner:2014gla,Denner:2016kdg}.
\item {\sc MadGraph5\_aMC@NLO}~\cite{Alwall:2014hca} (contact person: M. Zaro)
      is a meta-code, 
      automatically generating the source code to simulate any scattering process
      including NLO QCD corrections both at fixed order and including matching to parton showers. 
      It makes use of the FKS subtraction method~\cite{Frixione:1995ms, Frixione:1997np} 
      (automated in the module {\sc MadFKS}~\cite{Frederix:2009yq, Frederix:2016rdc}) 
      for regulating IR singularities. 
      The computations of one-loop amplitudes are carried out by switching dynamically 
      between two integral-reduction techniques, 
      OPP~\cite{Ossola:2006us} or Laurent-series expansion~\cite{Mastrolia:2012bu},
      and TIR~\cite{Passarino:1978jh,Davydychev:1991va,Denner:2005nn}. 
      These have been automated in the module {\sc MadLoop}~\cite{Hirschi:2011pa}, 
      which in turn exploits {\sc CutTools}~\cite{Ossola:2007ax}, 
      {\sc Ninja}~\cite{Peraro:2014cba, Hirschi:2016mdz}, or {\sc IREGI}~\cite{ShaoIREGI}, 
      together with an in-house implementation of the {\sc OpenLoops} optimisation~\cite{Cascioli:2011va}.\\
      The simulation of VBS at NLO-QCD accuracy can be performed 
      by issuing the following commands in the program interface:
      \begin{verbatim}
      > set complex_mass_scheme #1
      > import model loop_qcd_qed_sm_Gmu #2
      > generate p p > e+ ve mu+ vm j j QCD=0 [QCD] #3
      > output #4
      \end{verbatim}
      With these commands the complex-mass scheme is turned on {\tt \#1}, then the NLO-capable model is loaded {\tt \#2}\footnote{Despite
      the {\tt loop\_qcd\_qed\_sm\_Gmu} model also includes NLO counterterms for computing electroweak corrections, it is not yet possible to compute such corrections 
      with the current version of the code.}, finally the process code is generated {\tt \#3} (note the {\tt QCD=0} syntax to select the purely-electroweak process)
      and written to disk {\tt \#4}. Because of some internal limitations, which will be lifted in the future version capable of computing both QCD and EW corrections, 
      only loops with QCD-interacting particles are generated.
\item {\sc VBFNLO}~\cite{Arnold:2008rz, Arnold:2011wj, Baglio:2014uba} (contact person: M. Rauch)
      is a flexible parton-level Monte Carlo for processes with electroweak bosons. It
      allows the calculation of VBS processes at NLO QCD in the VBF
      approximation also including the $s$-channel tri-boson contribution,
      neglecting interferences between the two. Besides the SM, also anomalous
      couplings of the Higgs and gauge bosons can be simulated.
\item The {\sc Powheg-Box}~\cite{Alioli:2010xd,Frixione:2007vw} (contact person: A. Karlberg) 
      is a framework for matching NLO-QCD calculations with parton showers.
      It relies on the user providing the matrix elements and Born phase space, 
      but will automatically construct FKS 
      subtraction terms and the phase space for the real emission.
      For the VBS processes all matrix elements 
      are being provided by a previous version of {\sc VBFNLO}~\cite{Arnold:2008rz, Arnold:2011wj, Baglio:2014uba} 
      and hence the approximations used in the {\sc Powheg-Box} are similar to those used in {\sc VBFNLO}.
\item The program {\sc Recola+MoCaNLO} (contact person: M. Pellen)
      is composed of a flexible Monte Carlo program 
      dubbed \mbox{\sc MoCaNLO}~\cite{MoCaNLO} 
      and the general matrix element generator {\sc Recola} \cite{Actis:2012qn,Actis:2016mpe}.
      To numerically evaluate the one-loop scalar and tensor integrals, 
      {\sc Recola} relies on the {\sc Collier} library \cite{Denner:2014gla,Denner:2016kdg}.
      These tools have been successfully used for the computation of the full NLO corrections 
      for VBS~\cite{Biedermann:2016yds,Biedermann:2017bss}.
\item {\sc Whizard}~\cite{Moretti:2001zz,Kilian:2007gr} (contact person: V. Rothe)
      is a multi-purpose
      event generator with the LO matrix element generator {\sc O'Mega}. It
      provides FKS subtraction terms for any NLO process, while virtual matrix
      elements are provided externally by {\sc
      OpenLoops}~\cite{Cascioli:2011va} (alternatively, {\sc Recola}~\cite{Actis:2012qn,Actis:2016mpe}
      can be used as well). {\sc Whizard} allows to simulate a
      huge number of BSM models as well, in particular for new physics in
      the VBS channel in terms of both higher-dimensional operators as well as explicit
      resonances.
\end{itemize}

The complete comparison of the codes will be published in a separate work. Here, we present some preliminary results obtained at LO $\mathcal O (\alpha^6)$ and including
NLO QCD corrections at fixed-order $\mathcal O (\alpha^6\alpha_s)$, for the process ${\rm p}{\rm p}\to\mu^+\nu_\mu{\rm e}^+\nu_{\rm e}{\rm j}{\rm j}$.
In Table~\ref{tab:wg1_codes} the details of the various codes are reported. In particular, it is specified whether:
\begin{itemize}
    \item all $s$- and $t/u$-channel diagrams that lead to the considered final state are included;
    \item interferences between diagrams are included at LO;
    \item diagrams which do not feature two resonant vector bosons are included;
    \item the so-called non-factorisable (NF) QCD corrections, that is the corrections where (real or virtual) gluons are exchanged between different quark lines,
        are included;
    \item EW corrections to the $\mathcal O (\alpha^5\alpha_s)$ interference are included. These corrections are of the same order as the NLO QCD corrections to
        the  $\mathcal O (\alpha^6$) term.
\end{itemize}
%
%
\begin{table}
    \footnotesize
    \begin{tabularx}{\textwidth}{l|X|X|X|X|X}
        \multirow{1}{*}{Code} & \multicolumn{5}{c}{process content}\\ \cline{2-6}
                                        & $\mathcal O(\alpha^6)$ $|s|^2/$ $|t|^2/|u|^2$  &  $\mathcal O(\alpha^6)$ interf.  &  Non-res.  &  NF QCD  &  EW corr. to $\mathcal O(\alpha^5\alpha_s)$  \\
        \hline
        \hline
        {\sc POWHEG}    &  $t/u$  &  No   &  Yes            &  No        &  No  \\
        {\sc Recola}    &  Yes    &  Yes  &  Yes            &  Yes       &  Yes \\
        {\sc VBFNLO}    &  Yes    &  No   &  Yes            &  No        &  No  \\
        {\sc Bonsay}    &  $t/u$  &  No   &  Yes, No virt   &  No        &  No  \\
        {\sc MG5\_aMC}  &  Yes    &  Yes  &  Yes            &  No virt   &  No  \\
        {\sc Whizard}   &  Yes    &  Yes  &  Yes            &  No        &  No  \\  
    \end{tabularx}
    \caption{\label{tab:wg1_codes} Summary of the different properties of the codes employed in the comparison.
             Among them, the table details whether virtual corrections are included (virt), 
             as well as non-factorisable (NF) QCD correction are present.
            }
\end{table}

We simulate VBS production at the LHC, with a center-of-mass energy $\sqrt s = 13 \TeV$. We assume five massless flavours in the proton, and employ the NNPDF~3.0 parton 
density~\cite{Ball:2014uwa}
with NLO QCD evolution (the {\tt lhaid} in LHAPDF6~\cite{Buckley:2014ana} for this set is 260000) and strong coupling constant $\alpha_s( \MZ ) = 0.118$. Since
the employed PDF set has no photonic density, photon-induced processes are not considered. Initial-state collinear singularities are factorised with the  ${\overline{\rm MS}}$ 
scheme, consistently with what is done in NNPDF.\\
We use the following values for the mass and width of the massive particles:
\begin{alignat}{2}
                  \Mt   &=  173.21\GeV,       & \quad \quad \quad \Gt &= 0 \GeV,  \nonumber \\
                  \MZOS &=  91.1876\GeV,      & \quad \quad \quad \GZOS &= 2.4952\GeV,  \nonumber \\
                  \MWOS &=  80.385\GeV,       & \GWOS &= 2.085\GeV,  \nonumber \\
               M_{\rm H} &=  125.0\GeV,       &  \GH   &=  4.07 \times 10^{-3}\GeV.
\end{alignat}
The pole masses and widths of the W and Z~bosons are obtained from the measured on-shell (OS) values \cite{Bardin:1988xt} according to
\begin{equation}
 M_V = \frac{\MVOS}{\sqrt{1+(\GVOS/\MVOS)^2}},\qquad  
 \Gamma_V = \frac{\GVOS}{\sqrt{1+(\GVOS/\MVOS)^2}}.
\end{equation}
The EW coupling is renormalised in the $G_\mu$ scheme \cite{Denner:2000bj} where
\begin{equation}
    G_{\mu}    = 1.16637\times 10^{-5}\GeV^{-2}.
\end{equation}
The derived value of the EW coupling $\alpha$, corresponding to our choice of input parameters, is 
\begin{equation}
 \alpha = 7.555310522369 \times 10^{-3}. \\
\end{equation}
We employ the complex-mass scheme~\cite{Denner:1999gp,Denner:2005fg} to treat unstable intermediate particles in a gauge-invariant manner.\\

Cross sections and distributions are computed within the following VBS cuts 
inspired by experimental measurements \cite{Aad:2014zda,Aaboud:2016ffv,Khachatryan:2014sta,CMS:2017adb}: 
\begin{itemize}
    \item The two same-sign charged leptons are required to have
        \begin{align}
         \ptsub{\Pl} >  20\GeV,\qquad |y_{\Pl}| < 2.5, \qquad \Delta R_{\Pl\Pl}> 0.3\,.
        \end{align}
    \item The total missing transverse energy, computed from the vectorial sum of the transverse momenta of the two neutrinos in the event,
        is required to be
        \begin{align}
          \etsub{\text{miss}}=p_T^{miss} >  40\GeV\,.
        \end{align}
    \item QCD partons (quarks and gluons) are clustered together using the anti-$k_T$ algorithm~\cite{Cacciari:2008gp} with distance parameter $R=0.4$. Jets are required
        to have
        \begin{align}
         \ptsub{\Pj} >  30\GeV, \qquad |y_{\Pj}| < 4.5, \qquad \Delta R_{\Pj\Pl} > 0.3 \,.
        \end{align}
        On the two jets with largest transverse-momentum the following invariant-mass and rapidity-separation cuts are imposed
        \begin{align}
         m_{\Pj \Pj} >  500\GeV,\qquad |\Delta y_{\Pj \Pj}| > 2.5.
        \end{align}
    \item When EW corrections are computed, real photons and charged fermions are clustered together using the anti-$k_T$ algorithm with 
        radius parameter $R=0.1$. In this case, leptons and quarks mentioned above must be understood as {\it dressed fermions}. 
\end{itemize}

\begin{table}[h!]
    \centering
    \begin{tabular}{c|c}
        Code  &  $\sigma[\rm{fb}]$  \\
        \hline
        \hline
        {\sc Bonsay}  &  $1.5524 \pm 0.0002$ \\
        {\sc MG5\_aMC}&  $1.547 \pm 0.001$  \\ 
        {\sc POWHEG}  &  $1.5573 \pm 0.0003$ \\
        {\sc Recola+MoCaNLO}  &  $1.5503 \pm 0.0003$ \\
        {\sc VBFNLO}  &  $1.5538 \pm 0.0002$ \\
        {\sc Whizard}&  $ 1.5539 \pm 0.0004 $   
    \end{tabular}
    \caption{\label{tab:wg1_LOrates} Rates at LO accuracy within VBS cuts obtained with the different codes used in this comparison, 
    for the ${\rm p}{\rm p}\to\mu^+\nu_\mu{\rm e}^+\nu_{\rm e}{\rm j}{\rm j}$ process. 
    The quoted uncertainty corresponds to the integration error.}
\end{table}
\begin{figure}[h!]
   \centering
   \includegraphics[width=0.49\textwidth,angle=0,clip=true,trim={0.4cm 2.5cm 0.cm 1.cm}]{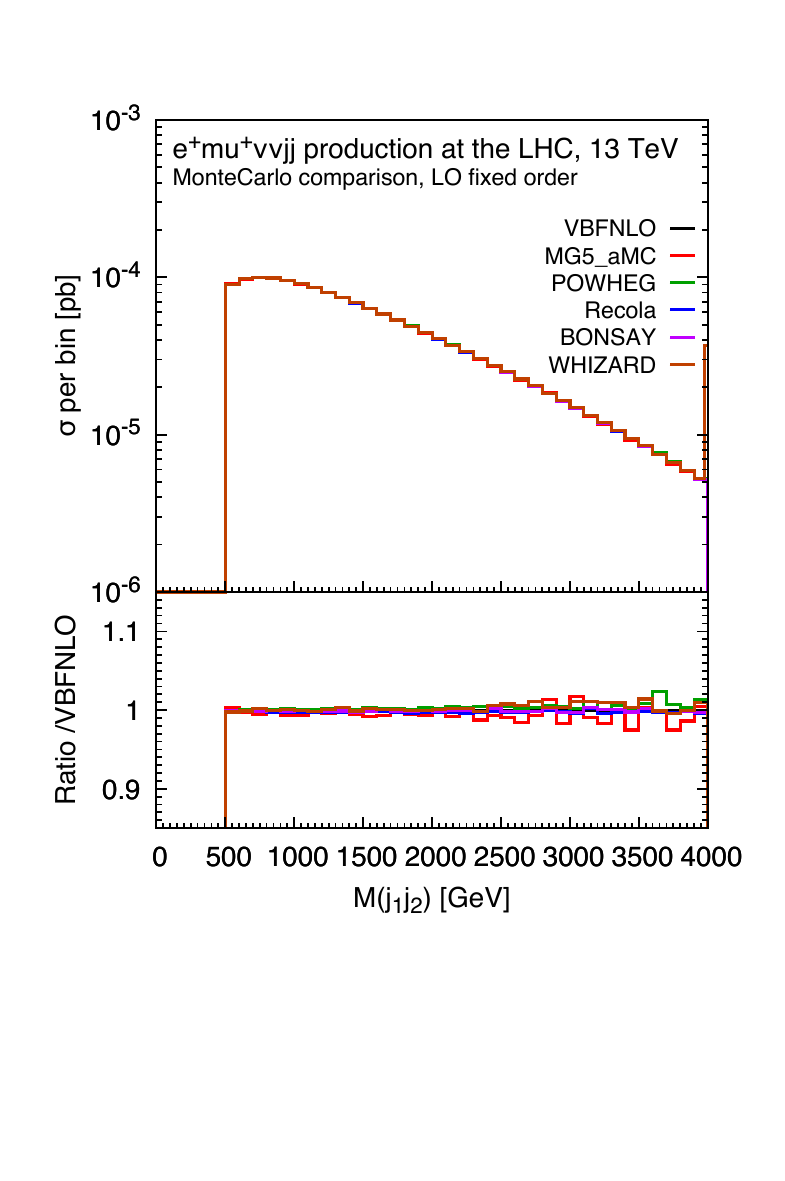}
   \includegraphics[width=0.49\textwidth,angle=0,clip=true,trim={0.4cm 2.5cm 0.cm 1.cm}]{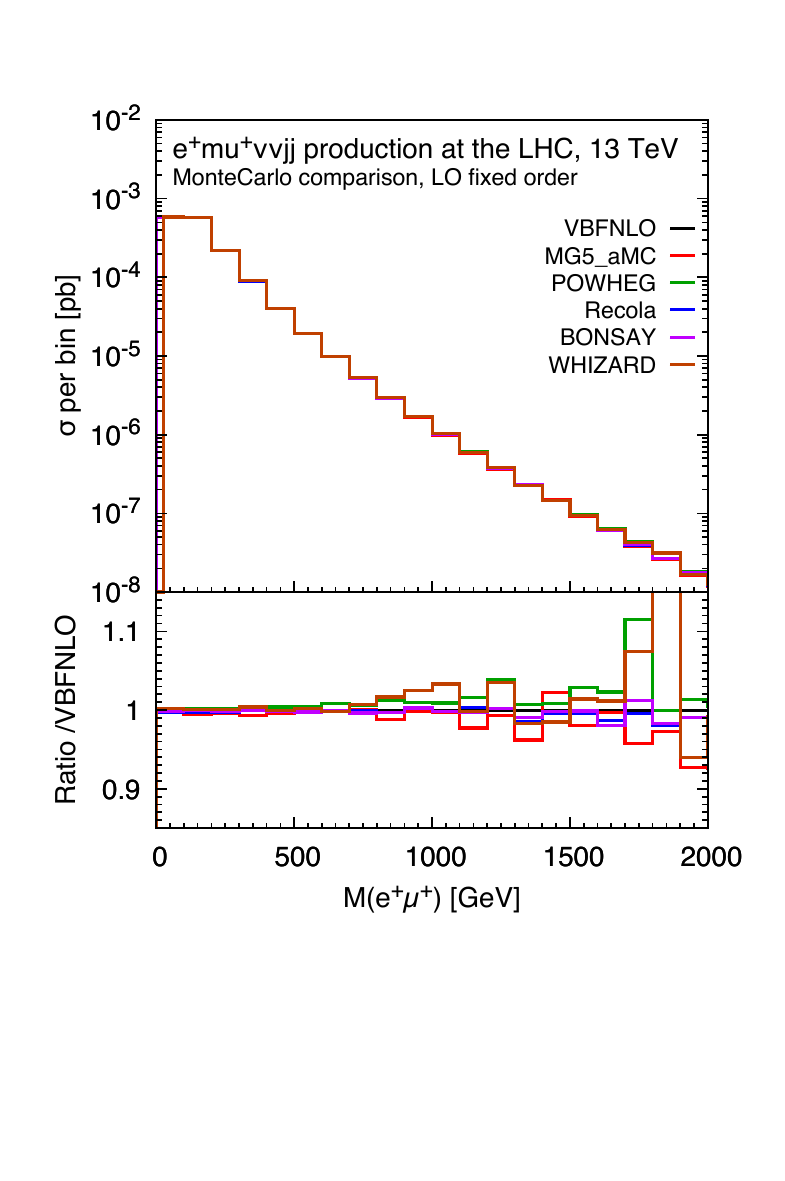}
\caption{\label{fig:wg1_mjj-llLO} Invariant-mass of the two tagging jets (left) and of the two leptons (right), at LO accuracy, 
computed with the different codes used in this comparison. The inset shows the ratio over {\sc VBFNLO}.
}
\end{figure}
\begin{table}[h!]
    \centering
    \begin{tabular}{c|c}
        Code  &  $\sigma[\rm{fb}]$  \\
        \hline
        \hline
        {\sc Bonsay}  &  $1.3366 \pm 0.0009$  \\
        {\sc MG5\_aMC}&  $1.318  \pm 0.003$  \\
        {\sc POWHEG}  &  $1.334 \pm 0.0003$  \\
        {\sc Recola+MoCaNLO}  &  $1.317 \pm 0.004 $ \\
        {\sc VBFNLO}  &  $1.3531 \pm 0.0003$  \\
    \end{tabular}
    \caption{\label{tab:wg1_NLOrates} Rates at NLO-QCD accuracy within VBS cuts obtained with the different codes used in this comparison, 
    for the ${\rm p}{\rm p}\to\mu^+\nu_\mu{\rm e}^+\nu_{\rm e}{\rm j}{\rm j}$ process. 
    The quoted uncertainty corresponds to the integration error.}
\end{table}
In Table~\ref{tab:wg1_LOrates} we report the total rates at LO accuracy obtained with the set-up described above, and in Figure~\ref{fig:wg1_mjj-llLO} we show the results
for the tagging jets (left) and lepton-pair (right) invariant-mass distributions. 
In both cases we show the absolute distributions in the main frame of the figures, 
while in the inset the ratio over {\sc VBFNLO} is displayed. 
For both observables we find a relatively good agreement among the various tools, 
which confirms the fact that contributions from $s$-channel diagrams 
as well as from non-resonant configurations are strongly suppressed in the fiducial region.
The same level of agreement is found for all other differential observables.
At NLO, rates show slightly larger discrepancies, as it can be observed in Table~\ref{tab:wg1_NLOrates}. 
This is most likely due to low dijet invariant-mass configurations, 
where $s$-channel diagrams and interferences are less suppressed than at LO, 
because of the presence of extra QCD radiation.

We conclude this section by recalling that the results presented must be regarded as preliminary.
In the coming months, this work will be enlarged to include comparison of predictions at NLO QCD matched to parton shower or with EW corrections, 
as well as to study the effect of changing 
the VBS cuts. The QCD-induced background will also be studied.

\subsection{Polarisation of vector bosons}\footnote{speaker: E. Maina}

Processes related to new physics could disturb the delicate balance which preserves unitarity
in VBS between longitudinally polarised vector bosons,
and lead to potentially large enhancements of the VBS rate, making it the ideal process for searches of 
deviations from the SM and hints of new physics. 
Developing methods which allow the separation of the different vector polarisations is, therefore, of primary
relevance.
A new technique has been proposed and applied to the scattering of two W bosons of opposite charge
\cite{Ballestrero:2017bxn}: the investigated process is $pp\rightarrow jje^-\mu^+\nu\nu$ at the LHC@13TeV, after VBS-like kinematic cuts.

The underlying formalism was established long time ago (see \emph{e.g.}\ References~\cite{Bern:2011ie,Stirling:2012zt}).
The polarisation tensor in the W propagator can be expressed in terms of polarisation vectors:
\begin{equation}
-g^{\mu\nu} + \frac{k^{\mu}k^{\nu}}{M^2} = \sum_{\lambda = 1}^4 \varepsilon^{\mu}_\lambda(k) 
\varepsilon^{\nu^*}_{\lambda}(k)\,\,.
\end{equation}

The decay amplitudes of the W depend on its polarisation.
In the rest frame of the $\ell\nu$ pair, they are:
\begin{equation}\label{eq:longamp}
\mathcal{M^D}_0 = ig\,\sqrt{2}E \,\sin\theta \,\,,\quad
\mathcal{M^D}_{R/L} = ig\,E \,(1 \pm \cos\theta)e^{\pm i\phi} \,\,,
\end{equation}
 where $0,R,L$ refer to the longitudinal, right, and left polarisations and $(\theta, \phi)$ are the charged 
 lepton angles relative to the boson direction.
Hence, each physical polarisation is uniquely associated with a specific angular distribution of the charged
lepton.    

Two issues, however, remain unresolved: 

\begin{itemize}
\item With few exceptions, electroweak boson production processes are described by amplitudes
including non resonant diagrams, which cannot be interpreted as
production times decay of any vector boson.
These diagrams are essential for gauge invariance and cannot be ignored.
For them, separating polarisations is simply unfeasible. 
\item Since the  W's are unstable particles, the decays of the individual 
polarisations interfere among themselves.
 \end{itemize} 

In order to define amplitudes with definite W polarisation, it is necessary to devise an accurate 
approximation to the full result that only involves double resonant diagrams.
Reference~\cite{Ballestrero:2017bxn} employed an on-shell projection (OSP) method,
similar to the procedure employed for
the calculation of EW radiative corrections to W$^+$W$^-$ production in
Reference~\cite{Denner:2000bj}.
 
It consists in substituting the momentum of the $\ell\nu$ pair  with a momentum on the W mass shell, 
while the denominator in the W propagator is left untouched. However,
this projection is not uniquely defined. In order to have an unambiguous
prescription one can choose to conserve:
the total four--momentum of the WW system (thus, also $M_{\rm WW}$ is conserved);
the direction of the two W bosons in the WW center of mass frame;
the angles of each charged lepton, in the corresponding W center of mass frame, relative to the boson
direction in the laboratory.
This procedure is gauge invariant. 

 If, for instance, one considers a polarised W$^-$
 boson and a non-polarised W$^+$ one, once all non double resonant diagrams have been dropped and the
 resonant ones have been projected,
the squared amplitude becomes:
\begin{equation}\label{eq:interfpol}
\underbrace{|\mathcal{M}|^2}_{\textrm{coherent sum}} = \underbrace{\sum_{\lambda}|
\mathcal{M}_{\lambda}|^2}_{\textrm{incoherent sum}} + \underbrace{\sum_{\lambda \neq \lambda'}
\mathcal{M}_{\lambda}^{ *}\mathcal{M}_{\lambda'}}_{\textrm{interference terms}}\,,
\end{equation}
where $\lambda$ is the W$^-$ polarisation.
In the absence of cuts on the final state leptons, the interference terms in Equation~\ref{eq:interfpol} cancel upon 
integration and the projected cross section is simply the sum of singly polarised cross sections. 
In the W center of mass frame the charged lepton angular distribution is: 
\begin{equation}
\frac{1}{\sigma(X)} \,\,\frac{d\sigma(\theta,X)}{d\cos\theta}
\ =\  \frac{3}{8} (1 + \cos\theta)^2 \,f_L
    + \frac{3}{8} (1 - \cos\theta)^2 \,f_R
    + \frac{3}{4} \sin^2\theta \, f_0 \,.
\label{eq:dcdist}
\end{equation}

The coherent sum of polarised amplitudes, obtained via direct computation (OSP method), differs by about 1\% from the exact cross section.
The differential distributions which do not depend on decay products of the
W bosons are equally well described.
Other variables, like the transverse momentum or the
angle $\phi$ of the electron, show sizeable differences  between the exact distributions and the projected 
ones.
The polarisation fractions obtained expanding the full result on the first three Legendre polynomials and through direct computation of singly 
polarised processes agree.
\begin{figure}[!tb]
\centering
\subfigure{\includegraphics[scale=0.5]{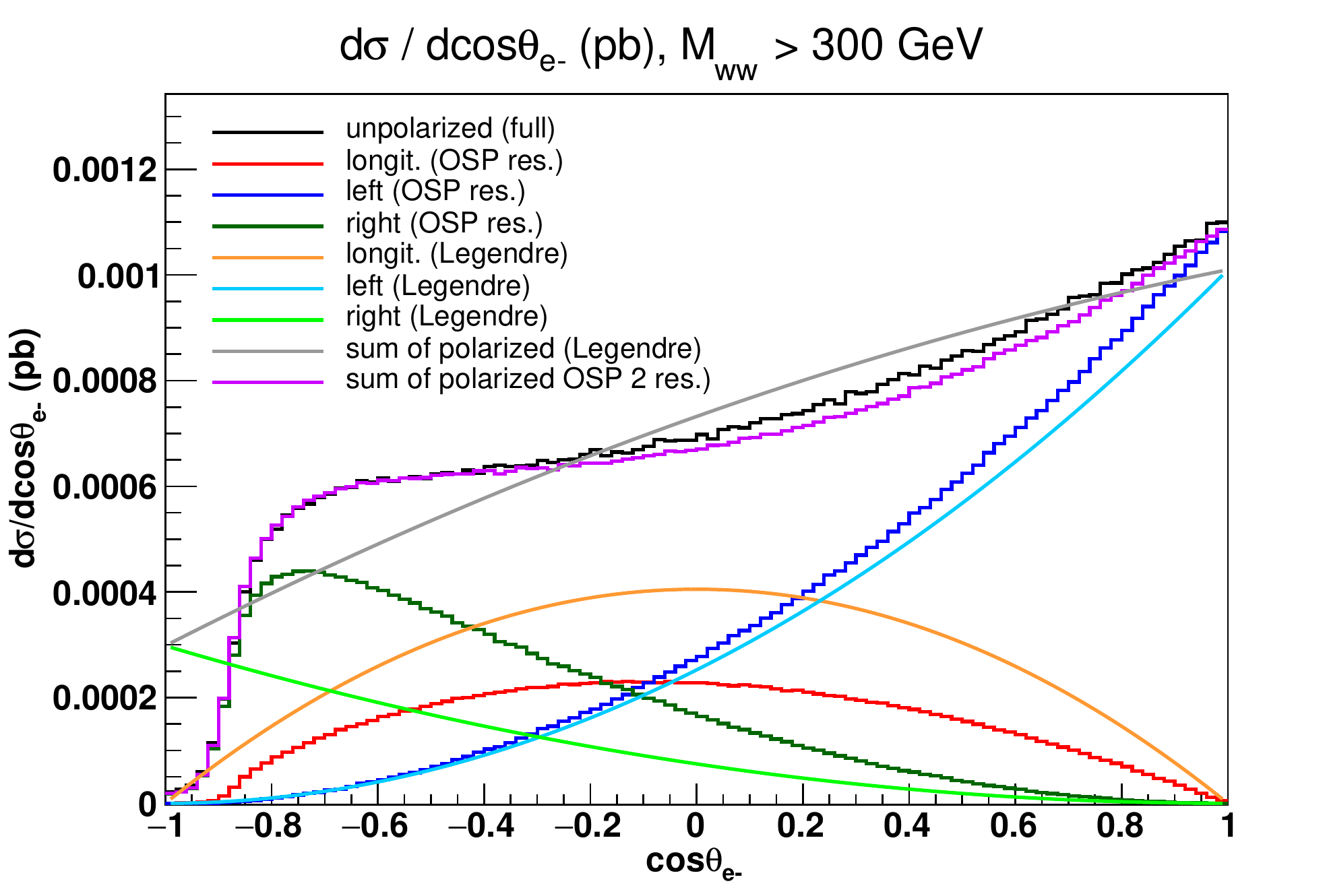}}
\qquad
\subfigure{\includegraphics[scale=0.5]{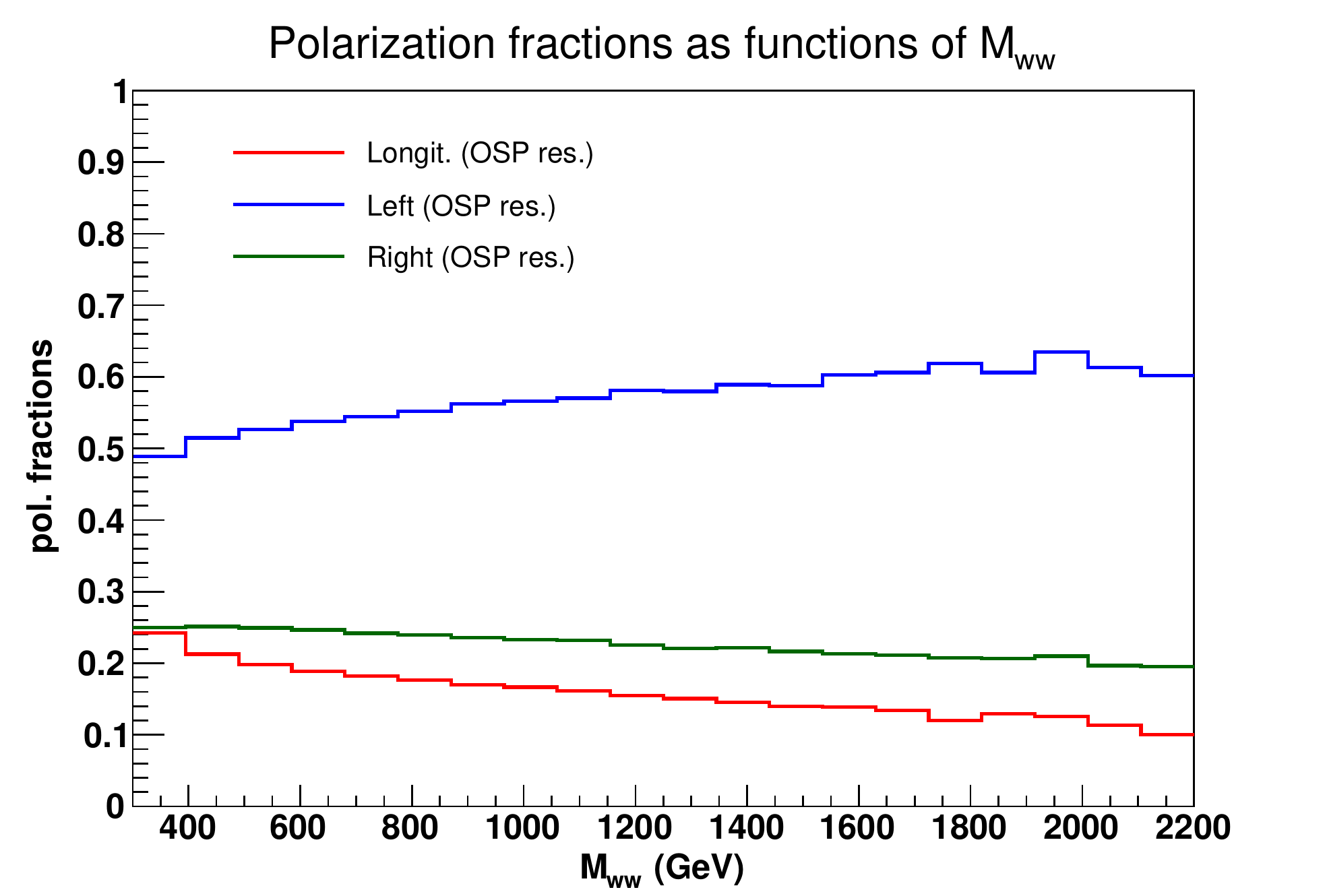}}
\caption{Distribution of $\cos\theta$ in the $W^-$ reference frame (top), polarisation fractions as functions of 
$M_{\rm WW}$ (bottom). Lepton cuts: $p_T^{e} > 20$ GeV, $|\eta^e| < 2.5$.}
\label{fig:distrib_thetal_polfrac_lepcut}
\end{figure} 

The approach proposed in Reference~\cite{Ballestrero:2017bxn} can be applied also when
standard acceptance cuts ($p_T^{\ell}>20 \,\,\mathrm{GeV}$ and
 $|\eta^{\ell}| < 2.5$) are imposed on both charged leptons.
In this case, the coherent sum of OSP polarised amplitudes differs from the full result by about 2\%.
The interference among polarisations is generally small. 
The distributions obtained from the incoherent sum of three OSP distributions agree  well with the full 
result for variables which do not depend on the W decay products.
The individual polarisations are not affected equally by the cuts. Typically, the cross section for right--
handed W's 
is reduced the most, followed by the cross section for longitudinally polarised W's. Left--handed W 
bosons seem to be the least sensitive to acceptance cuts.

The full angular distribution, Figure~\ref{fig:distrib_thetal_polfrac_lepcut}~(top),
is approximated within a few percent, over the full range, by the sum of the 
unpolarised distributions. 
The exact result, shown by the black histogram, however, is not of the form of Equation~\ref{eq:dcdist}.
This becomes clear expanding the full result on the first three Legendre polynomials
which yields the blue, green, and orange smooth 
curves in Figure~\ref{fig:distrib_thetal_polfrac_lepcut}: their sum (smooth grey curve) fails to describe the correct distribution,
thus the Legendre expansion cannot be applied when lepton cuts are imposed.\\
Computing the polarised cross section for different bins of WW invariant mass, we obtain the polarisation fractions
as functions of $M_{\rm WW}$, shown in the bottom panel of Figure~\ref{fig:distrib_thetal_polfrac_lepcut}: 
even in the presence of cuts, the fraction of longitudinally polarised W$^-$'s is well above 10\%.

The fact that the exact distribution is well described by the incoherent sum of the polarised differential 
distributions
allows for a measurement of the polarisation fractions, within a single model,
even in the presence of cuts on the charged
leptons, using Monte Carlo templates for the fit. 

This analysis demonstrates that it is possible to study the polarisation of massive gauge
bosons in a well defined set-up for VBS processes. The  method has been 
implemented in the code Phantom \cite{Ballestrero:2007xq}.

\subsection{Effective field theory for vector-boson scattering}\footnote{speaker: I. Brivio}

VBS processes represent a particularly interesting probe of new physics, as they give a unique access to the couplings of gauge bosons.
Without committing to a specific model, a convenient instrument for testing experimental data against the presence of BSM effects is that of effective field theories (EFTs).

In a EFT approach, the SM is assumed to be the low energy limit of an unknown UV completion, whose typical scale $\Lambda$ is well separated from the electroweak one.
In this scenario, the new physics sector is decoupled and its impact onto observables measured at $E\muchless\Lambda$ can be parametrised without specifying any property of the UV completion, by means of a Lagrangian that contains only the SM fields and respects the SM symmetries.
New physics effects are organised in a Taylor expansion in $E/\Lambda$, \emph{i.e.}\ they are encoded in an infinite series of gauge-invariant operators ordered by their canonical dimension.
This is often called SMEFT (SM EFT) Lagrangian and, neglecting lepton number violating terms, it reads
\begin{equation}
 \mathcal{L}_{\rm SMEFT} = \mathcal{L}_{\rm SM} + \frac{1}{\Lambda^2}\mathcal{L}_{\rm dim-6} + \frac{1}{\Lambda^4}\mathcal{L}_{\rm dim-8} + \dots \,,
\end{equation} 
with the dots standing for higher orders.
The SMEFT Lagrangian constitutes a convenient theoretical tool for probing the presence of new physics, as it provides the only systematic parameterisation of its effects that is both well-defined as a field theory and model-independent, in that it can be matched onto any UV completion compatible with the SM symmetries and field content.

One can restrict to leading deviations from the SM cutting the series at dimension 6 which reads
\begin{equation}
 \mathcal{L}_{\rm dim-6} = \sum_i C_i \mathcal{O}_i\, .
\end{equation} 
Here $\{\mathcal{O}_i\}$ is a set of gauge-invariant dimension-6 operators that form a complete basis and $\{C_i\}$ are the corresponding Wilson coefficients.
Any evidence for a non-zero Wilson coefficient would represent a smoking gun of new physics.
Further, knowing which terms are non-vanishing can allow to characterise the new physics states and help designing more effective direct search strategies.

A complete basis for dimension-6 terms contains 59 independent structures (+ their Hermitian conjugates) that in complete generality are associated to 2499 independent parameters~\cite{Alonso:2013hga}.
This number can be significantly reduced by assuming CP conservation and/or an approximate $U(3)^5$ flavour symmetry.
Choosing convenient kinematic cuts in the experimental measurements can also help to restrict the set of relevant operators.
Different basis choices for $\mathcal{L}_{\rm dim-6}$ have been proposed in the literature, 
that are related by equation-of-motion and integration-by-parts transformations. 
Despite containing different sets of operators 
(often distributing the effects differently among fermion and boson couplings), 
all the bases give equivalent parametrisations for physical $S$-matrix elements,
\emph{i.e.}\ once a complete process with stable external states is computed. 
The so-called Warsaw basis~\cite{Grzadkowski:2010es} is sometimes preferred, 
due to the fact that this was the first complete basis in the literature 
and that its renormalisation group evolution (RGE) and one-loop renormalisation 
are completely known~\cite{Jenkins:2013zja,Jenkins:2013wua,Alonso:2013hga,Grojean:2013kd,Alonso:2014zka,Ghezzi:2015vva}.

Assuming CP conservation and a $U(3)^5$ flavour symmetry, VBS processes receive corrections from 14 dimension-6 operators.
To keep the analysis as general as possible and to have a well-defined IR limit of a given underlying UV sector, these should be all considered simultaneously in the fit.
Setting a subset of the Wilson coefficients to zero cannot be done arbitrarily.
For example, this may spoil strong correlations hidden in the parametrisation and artificially remove blind 
directions\footnote{From a theoretical point of view, removing operators arbitrarily is problematic because a given basis is a minimal set in which a vast amount of redundant structures have already been systematically removed.
This means that each operator retained in the basis does not simply account for corrections to the couplings that it contains, but also to those contained in other structures related to it \emph{e.g.}\ by equations of motion, that have been removed~\cite{Passarino:2016saj, Passarino:2016pzb}.
This happens in a non-intuitive way, which is hard to control a posteriori.
For instance in the Warsaw basis some operators affecting triple gauge couplings (TGCs) are traded for a specific combination of fermionic + Higgs terms, which are apparently unrelated to the self-couplings of the gauge bosons.}.
In particular, including anomalous fermion couplings may have a significant impact on the analysis, despite the strong constraints imposed by LEP measurements (see \emph{e.g.}\ Reference~\cite{Baglio:2017bfe,Franceschini:2017xkh} for a recent study in the context of W$^+$ W$^-$ production at the LHC).
A reduction of the number of parameters may be necessary, nonetheless, for the technical feasibility of the analysis. In this case the removal of some (combination of) operators may be  very carefully considered in the future.

The possibility of extending the EFT analysis with dimension-8 operators has also been discussed, as these terms can introduce important decorrelation effects between triple and quartic gauge couplings.
Although this is an interesting avenue, exploring it in a consistent way is a challenging task due to the extremely large number of parameters involved 
(considering one fermion generation, there are 895 B-conserving independent operators at $d=8$, among which up to 86 can contribute to quartic gauge couplings (QGCs) and TGCs~\cite{Henning:2015alf}) 
and to the fact that a complete basis of dimension-8 operators is not available to date. 
Therefore it is advisable to defer this study to a later stage. A more compelling alternative is rather performing an analysis in the basis of the Higgs EFT (HEFT), 
for which complete bases have been presented in References~\cite{Buchalla:2013rka,Brivio:2016fzo} (see references therein for further theoretical details and previous phenomenological studies).
The HEFT differs from the SMEFT in that it is not 
constructed with the Higgs doublet, but rather embedding the Goldstone fields into a dimensionless matrix $\mathbf{U}=\exp(i\pi^a\sigma^a/v)$ (analogously to the pion fields in chiral perturbation theory) and treating the physical Higgs as a gauge singlet. The HEFT is more general than the SMEFT and it matches the low energy limit, for instance, of some theories with a strongly interacting electroweak symmetry breaking sector in the UV, such as composite Higgs models.
Such an analysis would be highly motivated as the scattering of longitudinal gauge bosons constitutes one of the best probes for UV scenarios matching the HEFT (see \emph{e.g.}\ References~\cite{Delgado:2013hxa,Delgado:2014jda} for recent studies), and they are among the observables that may allow to disentangle it from the SMEFT. The number of relevant Wilson coefficients for VBS in the HEFT (in the CP conserving, $U(3)^5$ symmetric limit) is about 30, which is larger than for the SMEFT but much lower than for including a complete dimension-8 set of operators, 
which makes this analysis an ideal follow-up to the SMEFT one.

One of the main points to be addressed in the EFT analysis is that of its validity: 
as mentioned above, 
adopting a dimension-6 parametrisation is theoretically justified 
only for $\Lambda$ sufficiently larger than the Higgs vev $v$.
Namely the impact of dimension-8 terms $\sim (E/\Lambda)^4$ should be roughly smaller than the experimental uncertainty.
When analysing experimental data, however, the cutoff scale $\Lambda$ is unknown and the actual energy $E$ exchanged in the process is often inaccessible too.
Extracting $E$ is particularly complex for VBS at the LHC, with various scales entering at different stages in the (sub-)process(es).
Thus the validity of the EFT cannot be established a priori: at best it can be verified a posteriori, checking that the energy range of the distributions used for the fit does not exceed the lower limit obtained for the cutoff.
Some methods of this kind have been discussed in the literature (see \emph{e.g.}\ References~\cite{Busoni:2013lha,Buchmueller:2013dya,Biekoetter:2014jwa,Englert:2014cva,Racco:2015dxa,Contino:2016jqw,Falkowski:2016cxu,Brivio:2017ije,Franceschini:2017xkh}) and could also be applied to VBS studies.
If a constraint is found to be incompatible with the validity of the EFT itself, it should be rejected.
Attention should be paid to the application of unitarisation methods, that are often employed to correct the divergences obtained in the kinematic distributions of Monte Carlo generated signals. Introducing a damping of the distribution tails, these techniques may alter the behaviour of the Taylor series in a way that does not reflect the correct behaviour of the EFT at high energies (which is indeed divergent where the expansion breaks down) and lead to an incorrect estimation of the constraints.

The first step of the EFT-VBS program is an accurate theoretical study of VBS in the SMEFT at dimension-6, which includes agreeing on a given parametrisation, evaluating the necessity of reducing the number of operators considered and testing the capabilities of available theoretical tools (Monte Carlo generators etc).
This will be conducted in parallel with a preliminary study of the experimental constraints that could be obtained. One of the primary goals of these studies, in which both theorists and experimentalists will participate, is to define an optimal way to report data (cross sections and differential distributions) that maximises the transparency and versatility of the results.
Finally, further avenues are worth exploring in subsequent stages, among which the analysis of the HEFT basis (and later on, if possible, of dimension-8 operators) and a comparison of the impact of VBS processes with that of other data sets, with the possibility of considering a combination of different measurements in the fit.


\section{WG2: Analysis techniques}
\label{WG2}

\subsection{Experimental Overview}\footnote{speaker: N. Lorenzo Martinez}
 
At the time of the workshop, 
a number of experimental results in VBS have been made available, 
all of them from the LHC experiments CMS and ATLAS (see Table~\ref{tab:wg2:expres}). 
The highlight among these results is a measurement from the CMS experiment in the W$^\pm$W$^\pm$ channel,  
which for the first time observes the VBS contribution in EW processes 
with a significance above $5\sigma$ (5.5 $\sigma$ observed)~\cite{CMS:2017adb}. 
It is interesting to notice that apart from this observation only two other evidences have been found, 
one for the VBS production in the Z$\gamma$ channel and one for the exclusive production $\gamma\gamma\rightarrow WW$.
The lack of evidences and observations of the VBS process 
is a consequence of its very small cross-section (order of 1~fb), 
but also of the large size of systematic uncertainties 
coming from background evaluation and the jet reconstruction in the forward region of detectors. 

\begin{table}[htb]
\centering
\begin{tabular}{l|c|c|c|c|c|c}
    Channel & \multicolumn{2}{c|}{$\sqrt{s}$} & \multicolumn{2}{c|}{Luminosity [fb$^{-1}$]} & \multicolumn{2}{c}{Observed (expected) significance} \\
    \hline
    \hline
                                 & ATLAS                 & CMS     & ATLAS & CMS  & ATLAS                                                             & CMS                                                  \\
    $Z(\ell\ell)\gamma$          & 8 TeV                 & 8 TeV   & 20.2  & 19.7 & 2.0$\sigma$ (1.8$\sigma$)\cite{Aaboud:2017pds}                    & 3.0$\sigma$ (2.1$\sigma$)\cite{Khachatryan:2017jub}  \\
    $Z(\nu\nu)\gamma$            & 8 TeV                 & --      & 20.2  & --   & Only aQGC lim. \cite{Aaboud:2017pds}                            & --                                                   \\
    $W^\pm W^\pm$                & 8 TeV                 & 8 TeV   & 20.3  & 19.4 & 3.6$\sigma$ (2.3$\sigma$)\cite{Aaboud:2016ffv},\cite{Aad:2014zda} & 2.0$\sigma$ (3.1$\sigma$)\cite{Khachatryan:2014sta}  \\
    $W^\pm W^\pm$                &  --                   & 13 TeV  & --    & 35.9 & --                                                                & 5.5$\sigma$ (5.7$\sigma$)\cite{CMS:2017adb}          \\
    $W(\ell\nu)\gamma$           &  --                   & 8 TeV   & --    & 19.7 & --                                                                & 2.7$\sigma$ (1.5$\sigma$) \cite{Khachatryan:2016vif} \\
    $Z(\ell\ell)Z(\ell\ell)$     &  --                   & 13 TeV  & --    & 35.9 & --                                                                & 2.7$\sigma$ (1.6$\sigma$) \cite{Sirunyan:2017fvv}    \\
    $W(\ell\nu)Z(\ell\ell)$      & 8 TeV                 & 8 TeV   & 20.2  & 19.4   & Only aQGC lim. \cite{Aad:2016ett}                             & N/A \cite{Khachatryan:2014sta}                       \\
    $W(\ell\nu)V(qq)$            & 8 TeV                 & --      & 20.2  & --     & Only aQGC lim. \cite{Aaboud:2016uuk}                           & --                                                  \\
    $\gamma\gamma\rightarrow WW$ & --                    & 7 TeV   & --    & 5.05 & --                                                                & $\sim 1\sigma$ \cite{Chatrchyan:2013akv}             \\
    $\gamma\gamma\rightarrow WW$ & 8 TeV                 & 7+8 TeV & 20.2  & 24.8 & 3.0$\sigma$ \cite{Aaboud:2016dkv}                                 & 3.4$\sigma$ (2.8$\sigma$) \cite{Khachatryan:2016mud} \\
  \end{tabular}
\caption{\label{tab:wg2:expres} 
         Summary of all published experimental results on VBS processes by final state 
         with the details on luminosity and energy at the center of mass $\sqrt{s}$ used for the measurements. 
         When available both expected and observed significances are provided. 
         Channels for which ``Only aQGC limits'' were studied are indicated in the significance column.
         }
\end{table}

\begin{figure}[h!]
    \begin{center}
    \includegraphics[width=\textwidth,scale=1]{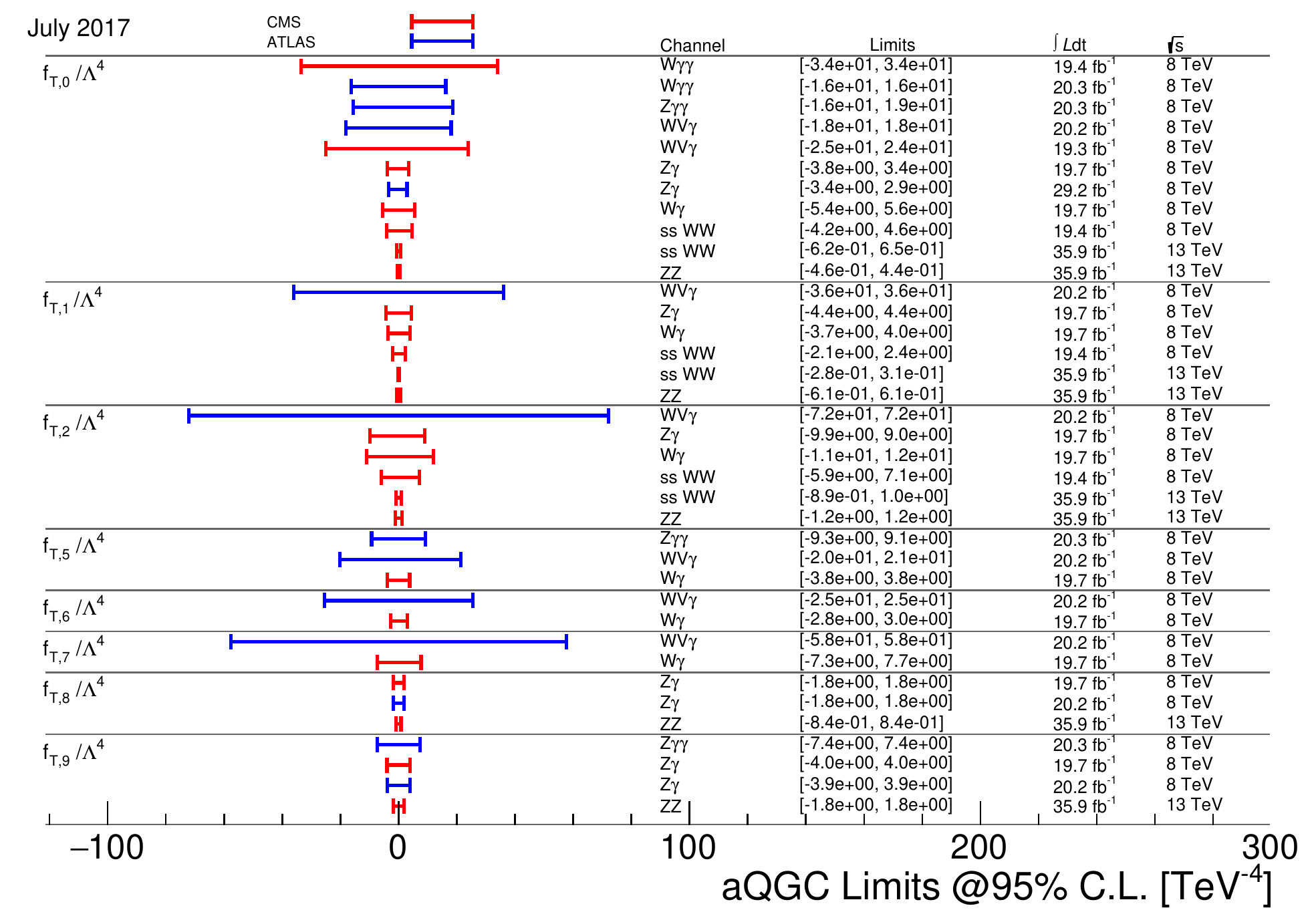}
    \end{center}
    \caption{Limits on dimension-8 mixed transverse and longitudinal 
             parameters $f_{M,i}$~\cite{CMSanCouplSumm}.}
    \label{fig:EFT}
\end{figure}

\begin{figure}[h!]
    \begin{center}
    \includegraphics[width=\textwidth,scale=1]{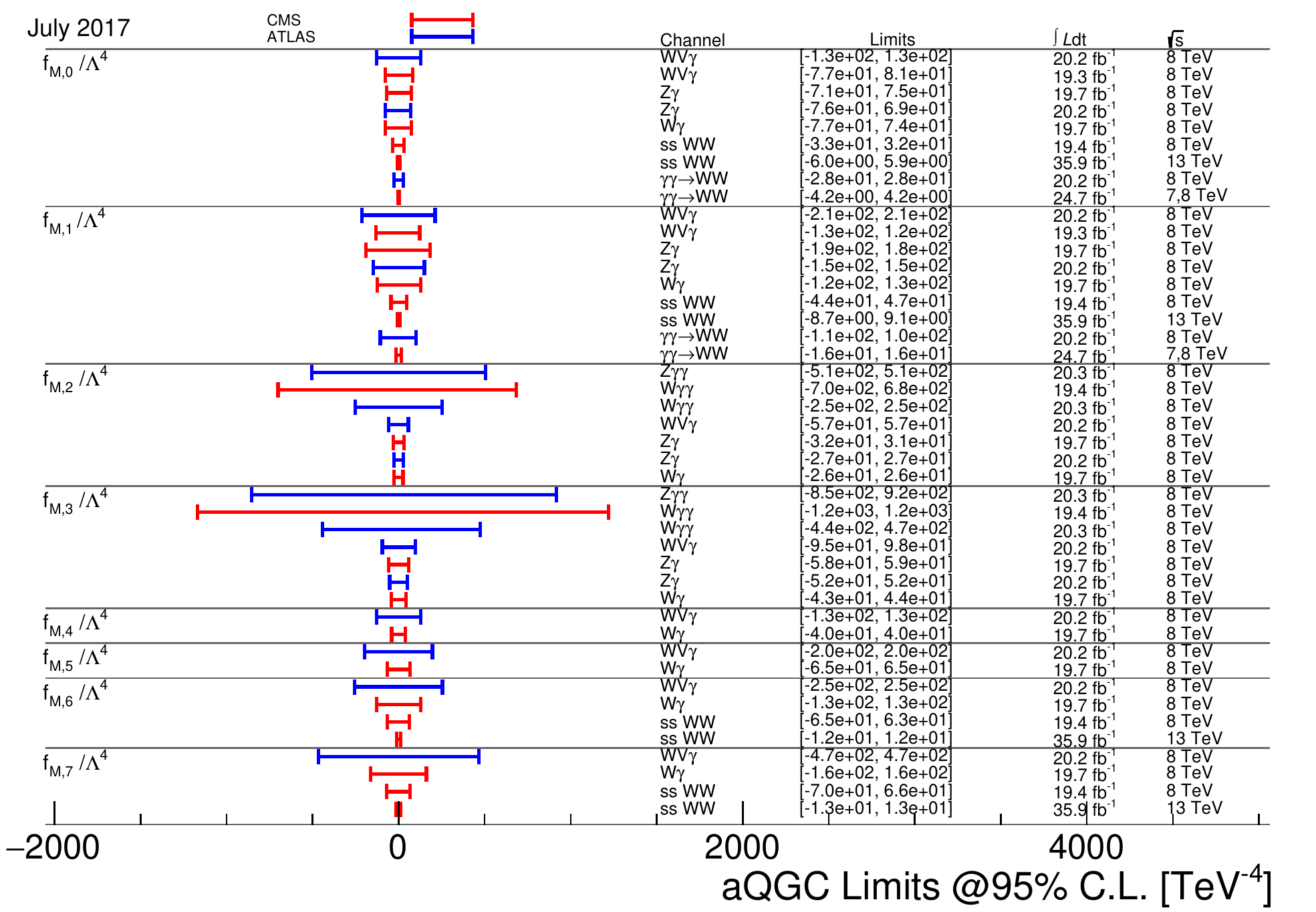}
    \end{center}
    \caption{Limits on dimension-8 transverse parameters $f_{T,i}$~\cite{CMSanCouplSumm}.}
    \label{fig:EFT2}
\end{figure}

By now, 
a large fraction of the different possible final state boson combinations 
have been studied by at least one experiment, 
with the notable absence of the $\gamma\gamma$ and W$^\pm$W$^\mp$ channels in the non-exclusive channel,  
which are much more complex to reach given the amount of experimental background 
associated to them. 
This wide coverage of channels has been shown to be very helpful when constraining aQGCs, 
as the different channels show varying sensitivity to different operators, 
as shown in Figures~\ref{fig:EFT} and~\ref{fig:EFT2}.
With the larger datasets collected during the second phase of exploitation of LHC (Run2), 
experiments will be able to reach more easily the VBS phase space 
and be able to study channels that are not accessible yet.

The presentation and interpretation of results in the VBS studies reviewed during the workshop 
show some notable differences among the various analyses and experiments.

The first difference is on the treatment of interference between electroweak and QCD amplitudes 
in the predictions for SM cross sections. 
Most of the time, 
the interference is derived from the {\sc MadGraph}~\cite{Alwall:2007fs} 
or {\sc Phantom}~\cite{Ballestrero:2007xq} generators, 
and is treated as systematic uncertainty on the signal yield, 
leading to 5-10\% uncertainties. 
In some cases, on the other hand, 
the interference is treated as a signal ($W^\pm W^\pm$ for the ATLAS experiment) 
or neglected when found to be too small (generally below 4\%). 
Such differences could complicate the combination of results.
Even if, 
with respect to the current level of experimental accuracy, 
the differences in treatment of interference effects are still small, 
with the continued data-taking at the LHC a common approach is desirable.

Another difference is the framework in which the aQGCs results are interpreted. 
There are mainly two approaches: one based on the CP-conserving dimension-8 EFT operators 
that maintains $SU(2)_L \times U(1)_Y$ gauge symmetry 
of the type $\phi/\lambda{}4$ 
(for the ATLAS experiment the Z$\gamma$ channel, 
 for the CMS one the Z$\gamma$, W$\gamma$, $W^\pm W^\pm$, and ZZ ones), 
the other one based on the $\alpha_4$ and $\alpha_5$ coefficients 
of the two linearly independent dimension-4 operators contributing to aQGCs 
(for the ATLAS experiment the WZ, $W^\pm W^\pm$, WV semileptonic channels). 
Even if the conversion between the two frameworks can be done, 
it can be better for quick comparison and combination to adopt an unique framework.

Finally, 
another difference lies in the treatment of unitarity issues
that can arise when the analysis sensitivity to potential aQGCs 
is not high enough to exclude aQGC values
within the energy range where the EFT can be considered unitary.
In the $\alpha_4$, $\alpha_5$ framework, 
the unitarisation condition is imposed with the so-called K-matrix method, 
within the {\sc WHIZARD}~\cite{Kilian:2007gr} generator. 
In the dimension-8 EFT operator framework, 
different approaches are used by ATLAS and CMS, respectively. 
ATLAS scales the spectra with analytical form factors of the type $f_i/(1+s/\Lambda_{FF^2})^n$, 
where $n=2$ and $\Lambda_{FF}$ is the cut-off scale. 
CMS, on its side, provides a validity bound, 
\emph{i.e.}\ the scattering energy at which the observed limit 
would violate the unitarity, 
derived with the {\sc VBFNLO}~\cite{Arnold:2008rz} generator. 
Some of the results published by the CMS collaboration show that at the LHC scales, 
already many limits are set in the unitarity unsafe region. 
Choosing different approaches 
could severely complicate the combination of different aQGCs limits measurements. 
Performing such combinations 
could substantially increase the statistical power of the total dataset 
and could help to break degeneracies between the effects of different operators 
which may affect a single channel in similar ways.
Providing recommendations to unify the treatments mentioned above is an important goal for WG2. 

Another lesson learned from the experimental review 
concerns the modeling of the main background 
producing the same final state via QCD interactions,
which is most of the time very important. 
To control its impact on the analyses, 
experiments use control regions (generally low dijet invariant mass) 
in which they constrain and verify the QCD background normalization and shape. 
While until now the precision is not enough to constrain these quantities, 
with more statistics it will become more relevant and QCD modeling issues 
could become one limiting factor.
At the same time, 
care should be taken in the signal definition as well,
because of the presence of the interference between the electroweak and QCD production
of the VBS final state.

The measurements are currently dominated by statistical uncertainty, 
due to the very low VBS cross-sections. 
Then generally follow the uncertainty on the jet energy scale and resolution, 
the uncertainty on background estimation and theory uncertainties (scales, parton distribution functions). 
The experimental uncertainties together with the previous point 
will be  important to mitigate in the future: this is one of the goals of WG2 and WG3.
 
Finally, 
it is interesting to notice the importance of the WV channel, 
where V is a W or Z boson decaying hadronically. 
Thanks to advanced jet substructure techniques, 
this channel brings the tightest constraint on EFT charged parameters, 
since the boosted topology allows to reach higher energy regimes. 
The WG3 activities will focus on such techniques as well.

\subsection{Common Selection Criteria}\footnote{speaker: X. Janssen}

In order to facilitate feasibility studies and similar forward looking analyses 
in a way that allows for a fair comparison between such studies, 
it is useful to define a common baseline selection for a fiducial phase space. 
This baseline criteria aim at selecting the two bosons and the two jets that are found 
in the final state of VBS processes, without entering yet in a VBS-enriched region (this will be a 
subject for later discussions in WG1, WG2 and WG3). 
A single definition cannot serve as the base for every study for the following reasons:
\begin{itemize}
\item Different VBS channels and different boson decay modes 
      may require notably different selection criteria.
\item The experiments don't always have the same geometrical acceptances and efficiencies in the final state objects identification and reconstruction
(depends on the calorimeter and muon systems),
\item The experiments will evolve due to hardware upgrades 
      planned for the high luminosity phase of the LHC, 
      necessitating different assumptions on the detector acceptance 
      depending on the integrated luminosity used for the forward looking study.
\end{itemize}

A simple example can be taken from studies of VBS channels with signals large enough 
to be amenable to a simple ``cut \& count'' style analysis, 
notably the W$^\pm$W$^\pm$ channel. 
Based on published results, 
a phase-space region close to CMS as well as ATLAS studies 
has been identified (see Table~\ref{tab:wg2:phasespace}). 
This phase space definition can be used for forward-looking comparison studies 
within the experiments reaching up to, 
but excluding the high luminosity phase of the LHC,
 when experiments will undergo upgrades that will improve their acceptance.

\begin{table}[htb]
\centering
\begin{tabular}{l|l|l|l|l}
             & electrons & muons & jets & photons \\
    \hline
    \hline
    $|\eta|$ & $<2.5$  & $<2.4$ & $<4.5$ & $<2.5$ \\
    $p_T^{lead.}$ & $>25$~GeV & $>25$~GeV &$>30$~GeV &$>25$~GeV\\
    $p_T^{sublead.}$ & $>15$~GeV & $>15$~GeV &&\\                            
  \end{tabular}  

\caption{\label{tab:wg2:phasespace} Proposed phase space for studies in the $W^+W^-$ channel.}
\end{table}

These numbers differ from the ones quoted in Section~\ref{sec:MCcomparison}, that were used only for 
a Monte Carlo comparison purpose of the same VBS process, and then did not need to follow exactly the experimental constraints 
(for example the current lepton trigger energy thresholds). 

Further extrapolation into the future suffers from the problem
that the design work for the envisioned detector upgrades is not entirely completed yet, 
so that work in this area is necessarily somewhat speculative. 
Nevertheless a few general conclusions can be drawn 
from existing detector design documents~\cite{Contardo:2020886, CMS-PAS-SMP-14-008, Collaboration:2257755, CERN-LHCC-2015-020, ATL-PHYS-PUB-2013-006, ATL-PHYS-PUB-2016-025}.
Both experiments plan to improve their trigger systems 
to keep the thresholds for lepton triggers at a similar level to current running conditions, 
so that $p_T$ thresholds should remain close to current ones. 
Both experiments also plan to extend coverage of their respective tracking detectors 
up to $|\eta|\sim 4$. 

For studies of channels with very low cross section,                                                                                                                
branching fraction or efficiency, for example the ZZ$\rightarrow 4\ell$ channel~\cite{Sirunyan:2017fvv},
the phase-space defined above is not suitable, and the analysis is performed in a more inclusive phase-space
(by lowering the threshold on the lepton transverse momentum).

\subsection{Prospects}\footnote{speaker: M. Kobel}

\begin{figure}[!h]
    \begin{center}
    \includegraphics[width=\textwidth,scale=1]{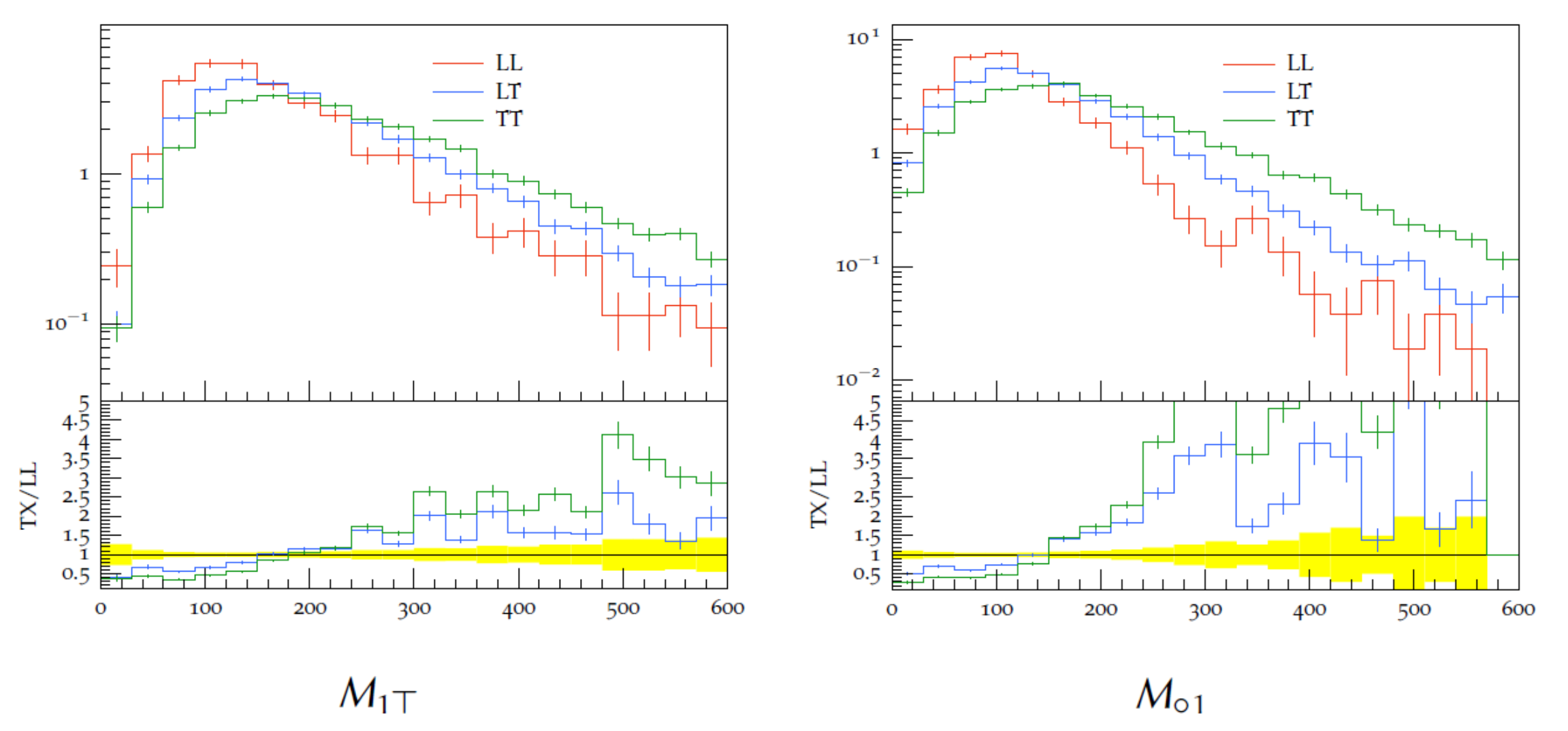}
    \end{center}
    \caption{Contributions from the different polarization states as a function of the mass-like variables $M_{1\top}$ on the left and the $M_{\circ 1}$ on the right~\cite{UlrikeMBI:2015}.}
    \label{fig:WW_mass-like}
\end{figure}

The presentation touched on several major topics interesting for future studies. 
First among these was the study of the final state bosons polarisation fractions. 
The scattering of longitudinal vector bosons violates unitarity 
in the absence of a standard model Higgs boson. 
By looking at the scattering of electroweak gauge bosons 
we will be probing the Higgs boson properties. 
At the LHC by the time of the workshop the $W^\pm W^\pm \rightarrow \ell \nu \ell \nu$ channel was the only one with observation of weak boson scattering. 
The polarisation is accessible through the angular distributions of the boson decay products in the boson rest frame, 
but in this particular channel the decay products include two neutrinos, 
which prevent a straight-forward reconstruction of the boson decay angular distribution.
Other channels, which allow for the reconstruction of these angular distributions, 
on the other hand, suffer from much larger backgrounds. 
Several potential approaches to address the issue of the missing information were presented.
One of them is to use a number of mass-like variables~\cite{Barr:2011,UlrikeMBI:2015} that have shown to be able to distinguish between the $W$ polarisations. Figure~\ref{fig:WW_mass-like} shows the contributions from the different polarization states as a function of two of this potential mass-like variables $M_{1\top}$\footnote{$M_{\circ 1}=\left(\left|\vec{p}_{T}(\ell_1)\right|+|\vec{p}_{T}(\ell_2)|+|\vec{p}_{Tmiss}|\right)^2-\left(|\vec{p}_{T}(\ell_1)|+|\vec{p}_{T}(\ell_2)|+|\vec{p}_{Tmiss}|\right)^2$} and the $M_{\circ 1}$\footnote{$M_{1\top} = \left(\sqrt{M^2_{\ell \ell}+ \vec{p}(\ell_1)+\vec{p}(\ell_2)} + |\vec{p}_{Tmiss}|\right)^2 - \left(\vec{p}(\ell_1)+\vec{p}(\ell_2)+\vec{p}_{Tmiss}\right)^2$}. 
The use of this variables may allow to
extract polarisation information even from final states 
with two neutrinos and might also be sensitive to new physics effects.
Another possibility to access the polarisation will be the use of a regression technique for pulling out the missing information~\cite{Baldi:2015}. 
In this sense it might be interesting to use a deep learning technique to output the true polarization values~\cite{Searcy:2015apa} using as input measurable quantities like leptons, jets and missing energy ($p_T$, $\eta$, $\phi$ etc.). 

\begin{figure}[!h]
    \begin{center}
    \includegraphics[width=0.7\textwidth,scale=1]{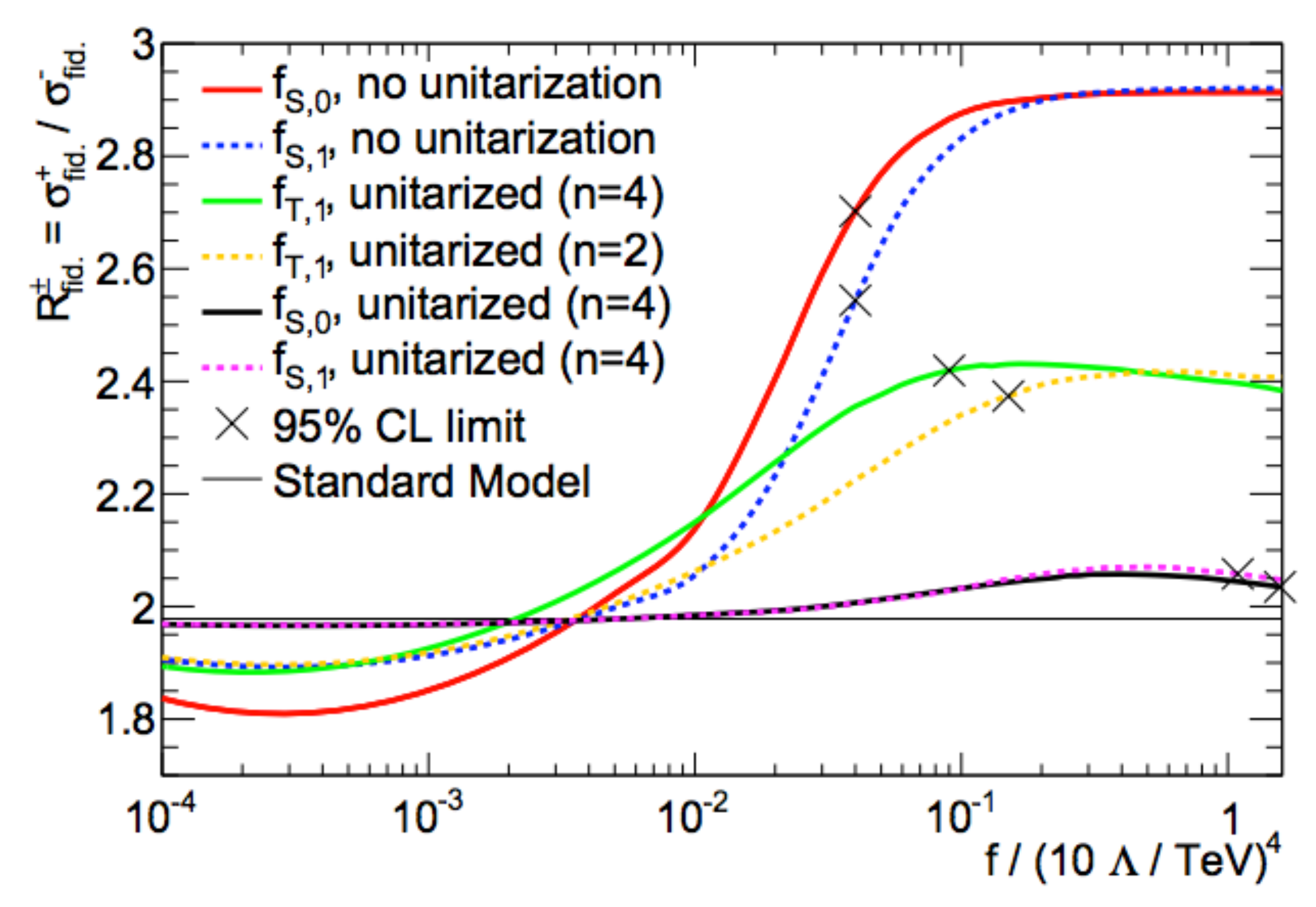}
    \end{center}
    \caption{Next-to-leading order fiducial cross section ratios of the electroweak
$W^{+}Zjj$ and $W^{-}Zjj$ production in the electroweak $WZjj$ fiducial phase space as a function
of different aQGC parameters for fully unitarized and the ununitarized processes~\cite{Anger:2014}.}
    \label{fig:WZChargeRatio}
\end{figure}

As a topic for future studies, 
the presentation discussed potential information to be gained 
by measuring the relative cross sections of different VBS channels, 
in particular channels related by charge symmetry. 
Beyond the naive expectations related to the valence quark content of the proton, 
the charge cross section ratio can provide useful constraints on the effects of the underlying parton density functions~\cite{Arrington:2012}.
It was shown as well that these ratios can provide sensitivity to BSM processes~\cite{Anger:2014}. 
As an example Figure~\ref{fig:WZChargeRatio} shows for the $W^{\pm} Zjj$ channel 
the charge cross section ratio using different unitarization prescriptions and different aQGC parameters. For increasing aQGC parameters the charge ratio increases up to a plateau separating from the SM prediction.

Additionally, 
the interpretation of experimental results 
in terms of BSM effects was discussed. 
While the EFT approach discussed above has the advantage of being generic 
and provides a complete description of a very wide range of BSM models, 
there is also a number of shortcomings. 
However, anomalous couplings of a size that will cause observable effects at low scales, 
will  commonly violate unitarity at high scales, 
so that ad-hoc approaches to unitarization are applied 
which introduce a new model-dependence: the dependence on the unitarization scheme.
Beyond this, 
there is the issue that the EFT is only valid in the approximation 
that the scale of the observed scattering is much smaller than the scale $\Lambda$. 
Empirical studies, 
comparing explicit resonance models to EFTs~\cite{Daniel:2014}, 
show that the amount by which $\Lambda$ 
has to exceed the described region of data can be substantial, 
to the point where visible effects that are accurately described by an EFT 
would correspond to theories so strongly coupled, 
that their treatment in perturbation theory would be questionable. 
A way to avoid these problems would be an interpretation 
using a simplified resonance parameterization~\cite{Alboteanu:2008,Reuter:2013}, 
which does not run into unitarity issues, 
but is by construction less general than the EFT approach.

\section{Experimental measurements}
\label{WG3}

\subsection{Large R jets and boosted object tagging in ATLAS}\footnote{speaker: C. F. Anders}

At the high center of mass energies of the Large Hadron Collider 
even the heaviest known SM particles can be observed with large transverse momenta, 
in the so called boosted topology. 
Boosted W, Z, and Higgs bosons, and top quarks that decay to quarks 
will have highly collimated decay products and can therefore be reconstructed in a single jet 
with large radius parameter $R$. 
In ATLAS jets are reconstructed with the anti-k$_t$~\cite{Cacciari:2008gp} algorithm, 
usually requiring a transverse momentum of $p_T>200$~GeV 
and a radius parameter of $R=1.0$,
and are trimmed~\cite{Krohn:2009th} with the parameters $R_{\mathrm{sub}}=$ 0.2 and $f_{\mathrm{cut}}=$ 5\%. 

\begin{figure}[!h]
\begin{centering}
\includegraphics[width=0.48\textwidth]{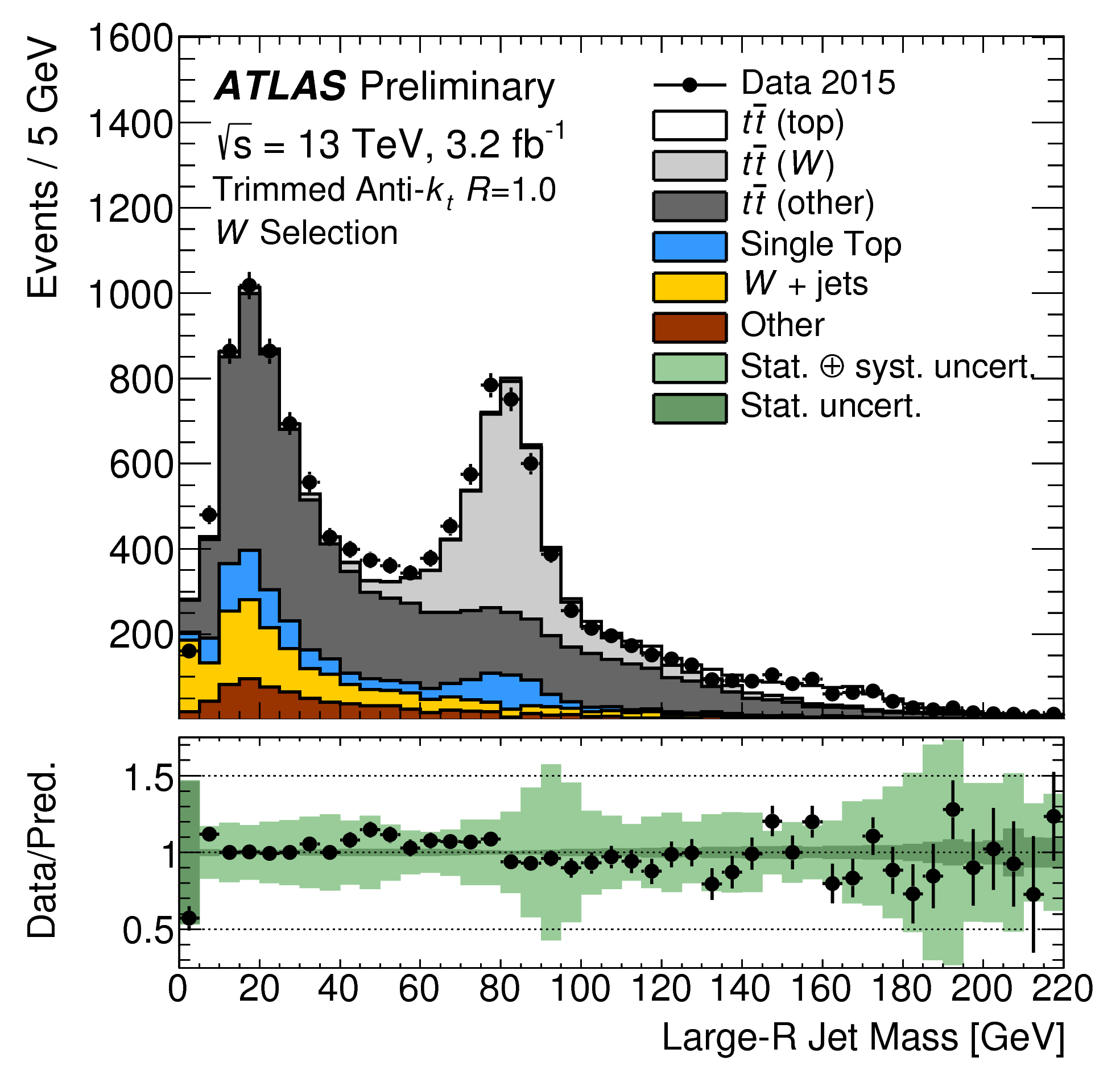}
\caption{Distribution of the calorimeter jet mass spectrum 
         for the leading-p$_T$ jet, in the ATLAS experiment,
         in 13~TeV data and MC simulation 
         using trimmed~\cite{Krohn:2009th} anti-k$_t$ R=1.0 jets,
         with trimming parameters $f_{\mathrm{cut}}=5$\% and $R_{\mathrm{sub}} = 0.2$ in lepton+jets events. 
         The large $R$ jets are required to have $p_T>200$~GeV~\cite{ATLASplots1}.}
\label{fig:jetmass_W}
\end{centering}
\end{figure}

To distinguish signal, 
\emph{e.g.}\ real W bosons, 
from QCD induced jet backgrounds one of the main observables is the jet mass, 
calculated from the jet constituents. 
In Figure~\ref{fig:jetmass_W} the distribution of the jet mass in data and simulation is shown 
after a lepton + jet selection that aims at selecting $t\bar{t}$ events. 
It shows a clear peak at the expected SM W boson mass. 
Using a further discriminating variable W boson taggers are built 
that have a fixed signal efficiency of 50\% and reduce the background by a factor of 50.

\begin{figure}[!h]
\begin{centering}
\includegraphics[width=0.48\textwidth]{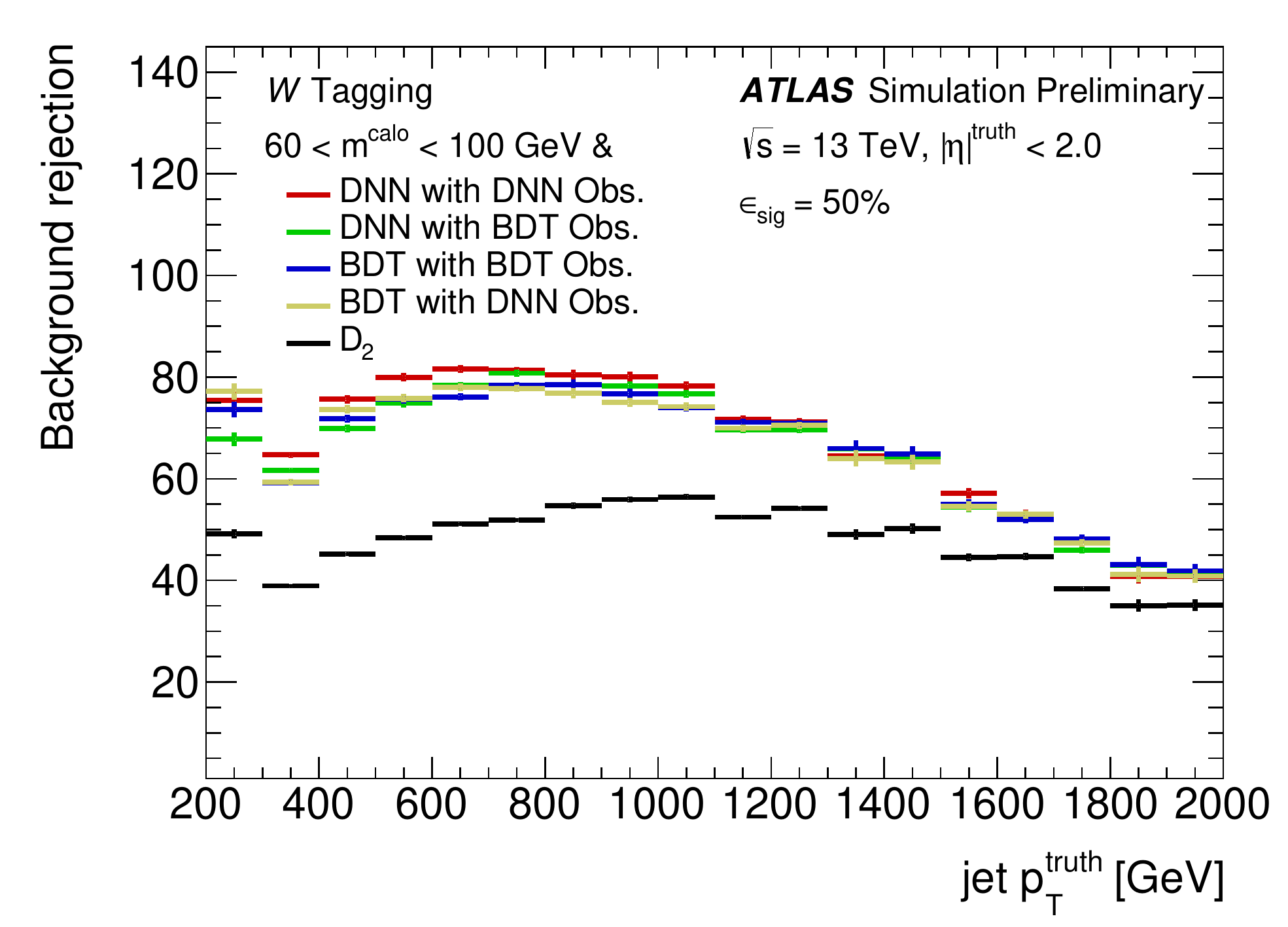}
\caption{Distributions showing comparison of the BDT and DNN taggers performance 
        to a simple W tagger~\cite{ATL-PHYS-PUB-2017-004} in the ATLAS experiment.}
\label{fig:MVA}
\end{centering}
\end{figure}

Recently, 
the possibilities of adding more information and exploiting multi-dimensional 
correlations have been explored by using Boosted Decision Trees (``BDT'') and Deep Neural Networks (``DNN'') 
to tag boosted hadronically decaying W bosons. 
Compared to simple 2-variable tagging approaches the multivariate ones perform better, 
as can be seen in Figure~\ref{fig:MVA}.

\subsection{Jet substructure techniques for VBS in CMS}\footnote{speaker: A. Hinzmann}

Jet identification techniques that make use of jet substructure information 
are important tools for the measurement of VBS.
A brief summary of the existing tools used by the CMS experiment and future prospects for the HL-LHC 
is given in the following.

To probe high WW, ZZ or WZ invariant masses, 
special jet identification techniques for W and Z bosons decaying to quarks are needed, 
since for high momentum W and Z bosons, 
the shower of hadrons 
originating from the quark anti-quark pair merges 
into a single large radius jet of particles~\cite{CMS-PAS-JME-16-003, CMS-PAS-JME-14-002, Khachatryan:2014vla}.
The maximum angular separation between the quark and anti-quark is given by $\Delta R_{q\bar{q}}=2 m / p_{T}$, 
where $m$ and $p_T$ are the mass and transverse momentum, respectively, of the W or Z boson.
For a W boson with $p_T=1$ TeV, 
an angular separation of  $\Delta R_{q\bar{q}}=0.2$ is expected, 
which is well below the typical jet size parameter of 0.4 used by CMS.
At even higher $p_T=3.5$ TeV, 
the angular distance $\Delta R_{q\bar{q}}=0.05$ between the decay products of a W boson 
is even smaller than the granularity of the hadron calorimeter of CMS 
with cell sizes of $\Delta \eta \times \Delta \phi = 0.087 \times 0.087$ in the barrel region of the detector.
CMS thus employs a particle-flow event algorithm~\cite{Sirunyan:2017ulk} to measure jet substructure, 
that reconstructs and identifies each individual particle with an optimized combination of information 
from the various elements of the CMS detector, 
benefiting from spatial and energy resolution of all sub-detectors.

\begin{figure}[htb]
\begin{center}
\includegraphics[width=.35\textwidth]{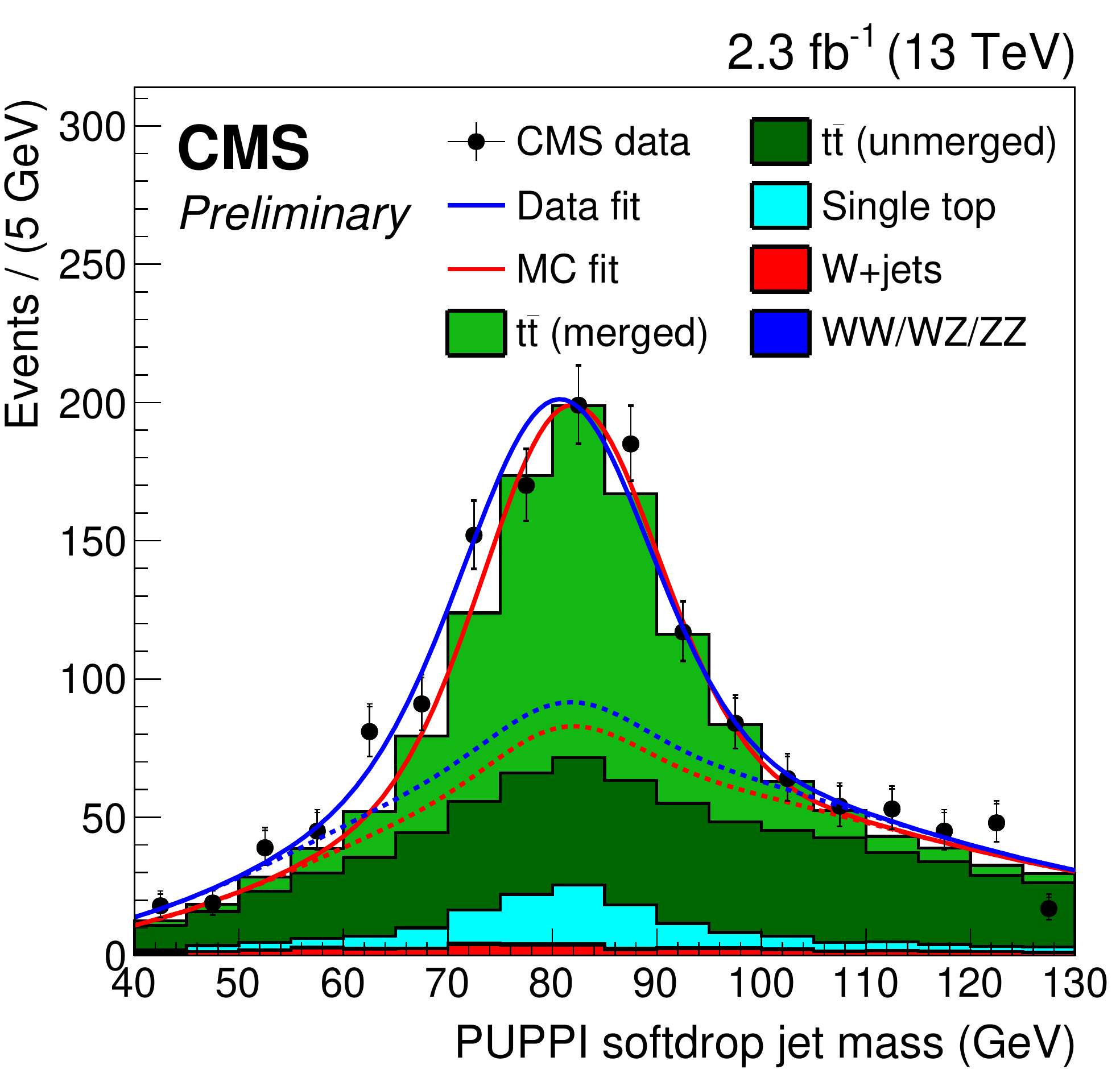}
\includegraphics[width=.48\textwidth]{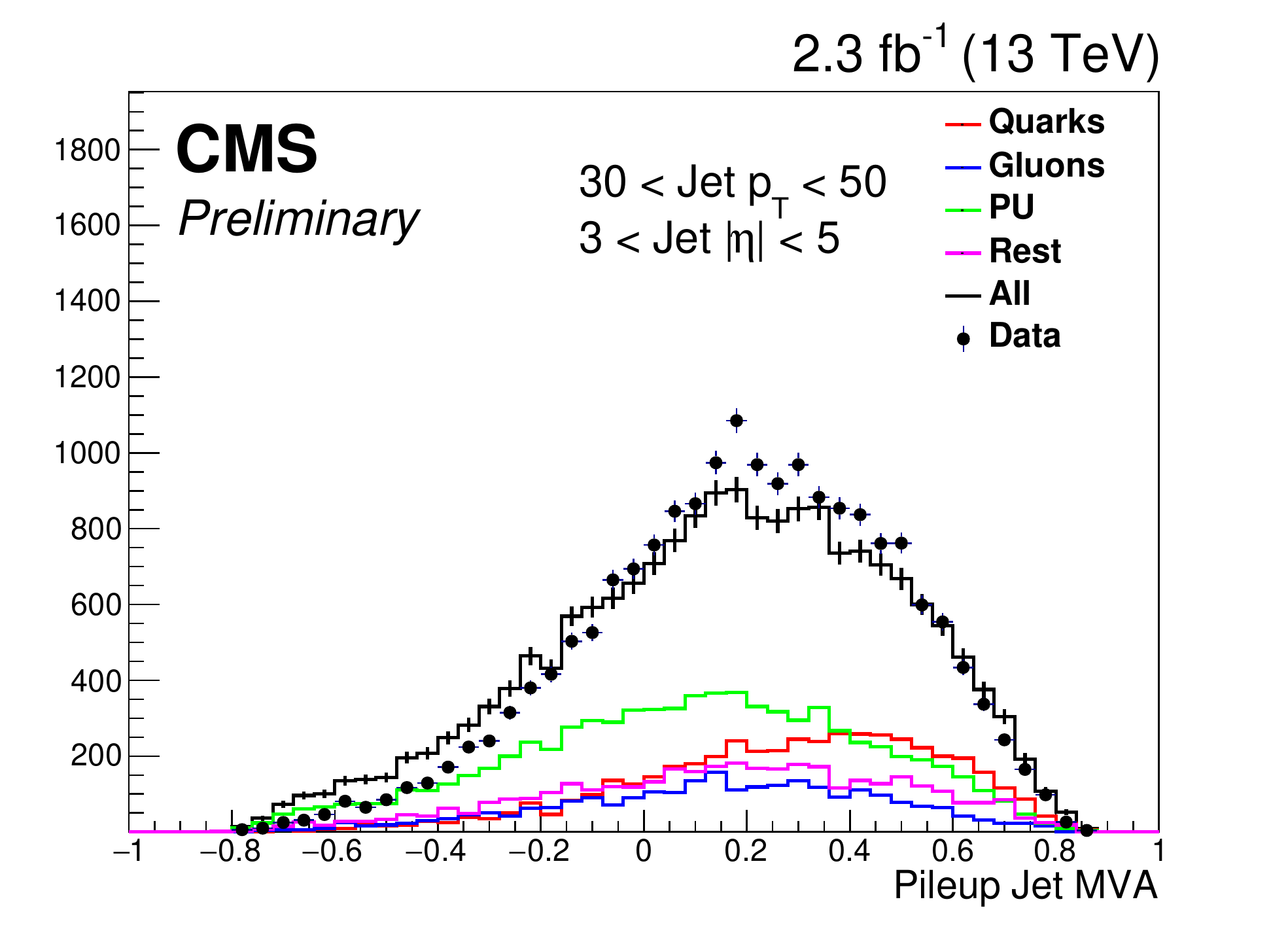}
\end{center}
\caption{(Left) Softdrop jet mass of boosted W bosons in data and simulated samples of top pair production 
         in the single lepton plus jets final state.
         (right) Pileup jet MVA discriminator in data and simulation for jets with $|\eta|>3.0$.
         The studies have been performed by the CMS collaboration~\cite{CMS-PAS-JME-16-003}.}
\label{fig:CMSsubstructure}
\end{figure}

Figure~\ref{fig:CMSsubstructure} (left) shows the main observable used by CMS 
to distinguish W and Z boson jets from quark and gluon initiated jets, the softdrop jet mass, 
which is the mass of the jet after iteratively removing soft radiation 
with the modified mass-drop algorithm~\cite{Dasgupta:2013ihk,Butterworth:2008iy}.
This ``softdrop'' algorithm~\cite{Larkoski:2014wba}
reduces the mass of quark and gluon jets and improves the mass resolution of W and Z boson jets.
In addition, 
the substructure of the jet is explored with an N-subjettiness~\cite{Thaler:2010tr} ratio 
that distinguishes the W and Z boson jets composed of two hard subjets from single quark and gluon jets.
With this combination of observables, 
a mistag rate of $\sim1$\% at an efficiency of 50\% is achieved 
for a broad range of jet transverse momenta from $p_T>200$ GeV up to at least 3.5 TeV.
The jet mass and substructure observables are calibrated 
using a data sample of top pair production in the single lepton final state containing high-$p_T$ W bosons, 
achieving uncertainties of the order of 1\% on jet mass scale and 10\% on jet mass resolution 
and jet substructure tagging efficiency, 
which increase at higher jet $p_T$ where simulation is used for extrapolation.
Particles from additional interactions happening in the same pp bunch crossing, 
called pileup interactions, 
can significantly distort these observables.
CMS thus employs dedicated particle-based pileup removal techniques that correct not only jet momenta, 
but also jet shape and substructure observables.
Charged particles that are identified by the tracking detector to originate from pileup interaction vertices 
are removed before jet clustering (this procedure is called charged hadron subtraction, {\sc
CHS}). 
For neutral particles a probability weight based on the distribution of surrounding particles 
following the pileup per particle identification ({\sc PUPPI}) algorithm~\cite{Bertolini:2014bba} 
is applied to the particle four momenta~\cite{CMS-PAS-JME-14-001}.
With these pileup suppression techniques, 
the performance of W and Z boson identification is constant up to at least 40 pileup interactions, 
and with the higher granularity tracking detector planned for the HL-LHC, 
performance is maintained up to 200 pileup interactions.
Notable for VBS studies, 
the longitudinal and transverse polarization of W boson jets can be separated using subjet information, 
yielding a mistag rate of $\sim$30\% at an efficiency of 50\%~\cite{Khachatryan:2014vla}.

Also the two forward jets from VBS require an analysis of their substructure, 
in order to suppress the background from (possibly overlapping) jets 
originating from pileup interactions~\cite{CMS-PAS-JME-16-003, CMS-PAS-JME-13-005}.
In a scenario of 25 pileup interactions (about half of the pileup expected in Run II of the LHC), 
without pileup mitigation $\sim$50\% of VBS selected forward jets come from pileup.
Since for $|\eta|>3.0$ no tracking information is available to suppress pileup particles 
and also jet shape differences are difficult to resolve due to coarse calorimeter granularity 
($\Delta \eta \times \Delta \phi = 0.175 \times 0.175$), 
a multivariate analysis (MVA) is needed to distinguish quark jets from pileup and gluon background.
Figure~\ref{fig:CMSsubstructure} (right) shows the pileup jet MVA discriminator 
that allows to achieve a pileup mistag rate of $\sim$30\% at a quark efficiency of 50\%.
At the HL-LHC, 
for which CMS tracking and vertex identification will be extended from $|\eta|<2.5$ to $|\eta|<4$ 
and a high granularity endcap calorimeter will be installed in the region $1.5<|\eta|<3$, 
VBS jet identification can rely on the much more powerful {\sc CHS} and {\sc PUPPI} pileup rejection techniques.

\cleardoublepage
\phantomsection
\addcontentsline{toc}{part}{Introduction}

\section{Summary}
\label{chapter:summary}

The VBSCan COST Action aims at a consistent and coordinated study of VBS
from the phenomenological and experimental points of view, 
gathering all the interested parties in the high-energy physics community,
together with experts of data mining techniques.
In fact, the complex struture of the process calls for a joint effort 
to exploit at best the data sets that will be delivered by the LHC
in the following years.
This will require a close interaction between the experimental community and the theory one,
as well as the deployment of the most advanced data analysis techniques
to maximise the reach of the ATLAS and CMS collaborations.
This document contains the summary of the kickoff meeting of the VBSCan Action,
happened in June 2017, 
and paves the way for the activities of the Network.

\phantomsection
\addcontentsline{toc}{part}{Acknowledgements} 

\section*{Acknowledgements}

The authors would like to acknowledge the contribution of the COST Action CA16108.
We would like to thank the Split-FESB team for the great hospitality in Split.
Kristin Lohwasser, Hannes Mildner, and Philip Sommer are supported 
by the European Union Horizon 2020 research and innovation programme under ERC grant agreement No. 715871.
Nigel Glover, Raquel Gomez-Ambrosio, Giulia Gonella, Markus Schumacher 
acknowledge the support of the Research Executive Agency (REA) of the European Union 
under the Grant Agreement PITN-GA-2012-316704 (``HiggsTools'').
Stefan Dittmaier and Christopher Schwan acknowledge support by the state of Baden-W\"urttemberg 
through bwHPC and the DFG through grant no. INST 39/963-1 FUGG and grant DI 784/3.
Benedikt Biedermann, Ansgar Denner, and Mathieu Pellen acknowledge financial support 
by the German Federal Ministry for Education and Research (BMBF) under contract no. 05H15WWCA1 
and the German Research Foundation (DFG) under reference number DE 623/6-1.
Andreas Hinzmann gratefully acknowledges funding in the Emmy-Noether program (HI 1952/1-1) 
of the German Research Foundation DFG.
Ilaria Brivio and Michael Trott acknowledge support from the Villum Foundation, 
NBIA, the Discovery Center at Copenhagen University 
and the Danish National Science Foundation (DNRF91).
Qiang Li is supported in part by the National Natural Science Foundation of China, 
under Grants No. 11475190, No. 11661141008 and No. 11575005,  
by the CAS Center for Excellence in Particle Physics (CCEPP).
Jan Kalinowski acknowledges support from the Polish National Science Center HARMONIA project 
under contract UMO-2015/18/M/ST2/00518 (2016-2019). 
The work of Barbara J\"ager is supported in part part by the Institutional Strategy 
of the University of T\"ubingen (DFG, ZUK 63) 
and in part by the BMBF under contract number 05H2015. 
Alexander Karlberg acknowledges financial support 
by the Swiss National Science Foundation (SNF) under contract 200020-175595. 
The work of Marc-André Pleier was supported by the DOE Contract No. DE-SC0012704.



\renewcommand\leftmark{References}
\renewcommand\rightmark{References}

\bibliographystyle{./StyleFilesMacros/atlasnote}
\cleardoublepage
\phantomsection
\addcontentsline{toc}{part}{References}
\bibliography{SplitReport}


\end{document}